\documentclass[suppldata]{interact}
\usepackage[T1]{fontenc}

\usepackage{epstopdf}
\usepackage[caption=false]{subfig}

\usepackage[numbers,sort&compress]{natbib}
\bibpunct[, ]{[}{]}{,}{n}{,}{,}

\theoremstyle{plain}

\theoremstyle{definition}

\theoremstyle{remark}

\usepackage{url}

\usepackage{caption}
\usepackage{placeins}
\usepackage{makecell}
\usepackage{adjustbox}
\usepackage{graphicx}
\usepackage{tikz}
\usetikzlibrary{arrows.meta,positioning,calc}
\graphicspath{{./}}
\usepackage{booktabs}
\usepackage{lineno}
\usepackage[utf8]{inputenc}
\makeatletter
\def\abx@aux@refcontext#1{}%
\def\abx@aux@defaultrefcontext#1#2#3{}%
\def\abx@aux@cite#1#2{}%
\def\abx@aux@nocite#1#2{}%
\def\abx@aux@nociteall{}%
\def\abx@aux@segm#1#2#3{}%
\def\abx@aux@page#1#2{}%
\def\abx@aux@fnpage#1#2{}%
\def\abx@aux@refsection#1#2{}%
\def\abx@aux@category#1#2{}%
\def\abx@aux@biblist#1{}%
\def\abx@aux@defaultlabelprefix#1#2#3{}%
\def\abx@aux@number#1#2#3#4#5{}%
\def\abx@aux@backref#1#2#3#4#5{}%
\def\abx@aux@count#1#2{}%
\def\abx@aux@fncount#1#2{}%
\def\abx@aux@locallabelwidth#1#2#3{}%
\def\abx@aux@read@bbl@mdfivesum#1{}%
\def\abx@aux@read@bblrerun{}%
\def\abx@aux@lbx@loadrequest#1{}%
\makeatother
\begin{document}

\title{Racing to the Starting Line: Measuring Opportunity and Improvement in Cross Country}

\author{%
\name{Jonathan A. Karr Jr\textsuperscript{a,b,*}\thanks{CONTACT Jonathan A. Karr Jr. Email: jkarr@nd.edu; nationalrunningclubdatabase@gmail.com},
Ryan M. Fryer\textsuperscript{a,c,*}\thanks{Ryan M. Fryer. Email: jpv8cu@virginia.edu},
and Nitesh V. Chawla\textsuperscript{a}\thanks{Nitesh V. Chawla. Email: nchawla@nd.edu}}
\affil{\textsuperscript{a}University of Notre Dame, Notre Dame, IN, USA;
\textsuperscript{b}National Running Club Database;
\textsuperscript{c}University of Virginia, Charlottesville, VA, USA;
\textsuperscript{*}These authors contributed equally to this work.}
}

\maketitle

\begin{abstract}
Collegiate cross country programs set race schedules without large comparable datasets across courses and conditions. Using the comprehensive-era National Running Club Database (NRCD; 23{,}360 results; 7{,}056 athletes; 2023--2025), where course and weather coverage exceed 99\%, we ask what public results can identify about seasonal improvement, measurement, and racing opportunity. Hierarchical models of Standardized times recover pooled within-season improvement of about $-5.7$\,s/week for men and $-5.2$ for women, but reading those slopes as fitness requires residual meet difficulty not to track the calendar: crossed meet effects can erase the week slope, while repeating-venue effects recover a clearer men's estimate under a weaker assumption. After the opener, race-1-only forecasts of who will improve are near noise (held-out $R^2\approx 0$). Weather standardization remains a useful convention (Converted Only overstates first-to-last gains by about $15$--$21$\,s relative to Standardized), and prospectively trained venue factors modestly help where prior course data exist. Successful programs race more and place higher at nationals, but that may not be why they win: the schedule--placement link is between programs and is entangled with roster size, not evidence that changing one team's race count raises nationals place.
\end{abstract}

\begin{keywords}
cross country, course adjustment, athlete progression, observational inference
\end{keywords}

Code: \url{https://github.com/National-Running-Club-Database/nrcd_xc_paper}
\newpage

\section{Introduction}

How can a cross country program compare improvement across unlike courses and decide who needs more racing opportunities? Large public performance datasets were scarce before the National Running Club Database (NRCD) \citep{karr2026nrcd,nrcddataset2026}. Its resource paper supplies the data and Converted Only / Standardized times; here we ask what those times can identify about seasonal improvement and program structure.

Collegiate cross country comprises teams from governing bodies including the NCAA, NAIA, and NJCAA. Results are hosted on websites such as Athletic.net \cite{athleticnet}, MileSplit \cite{milesplit}, and TFRRS \cite{tfrrs}, but these platforms do not provide bulk relational exports, so prior collegiate analyses were often limited to hand-curated samples of roughly hundreds of records \cite{millettpredicting}. Sports science has also underrepresented women \cite{james2023underrepresentation, smith2022auditing}, which weakens the external validity of scheduling and improvement guidelines derived from men-only studies.

We study collegiate club athletes competing through National Intercollegiate Running Club Association programs \cite{nirca}. These programs span a wide ability range and face the same basic scheduling problem as varsity teams. We analyze only comprehensive-era Cross Country (23{,}360 results; 7{,}056 athletes; 2023--2025), where course and weather coverage exceed 99\%; the earlier historical era is too incomplete for the same comparisons.

Fair comparison across meets requires adjustment. Converted Only corrects distance to 8{,}000\,m (men) / 6{,}000\,m (women); Standardized additionally applies the adopted temperature--dew-point, grade, and altitude rules \citep{karr2026nrcd}. We keep Standardized primary, test its weakly identified heat-rule parameters, and prospectively evaluate repeated-venue factors. The central identifying assumption is stronger than high metadata coverage: \emph{after Standardization, residual meet difficulty must not remain correlated with calendar week}. We then ask what seasonal improvement is identified under that assumption and what changes when meet effects are introduced.

\noindent\textbf{Research questions:}
\begin{itemize}
\item \textbf{RQ1:} How much do collegiate club runners improve within a cross country season after accounting for course and weather, and which early-season features predict end-of-season improvement rate?
\item \textbf{RQ2:} Do athletes with stable multi-season participation continue to improve across consecutive seasons?
\item \textbf{RQ3:} Where does gender imbalance in club cross country arise, and does the race-count/improvement association differ by sex?
\item \textbf{RQ4:} How is team racing opportunity associated with nationals placement, and does that association operate within or between programs?
\end{itemize}
Gender-stratified fits are the primary RQ1 and RQ2 models. RQ3 asks whether the observed participation gap reflects fewer recorded competitors or different racing intensity among those competitors, and whether the race-count association with improvement differs by sex. RQ4 separates the cross-sectional program association from within-program year-to-year changes. The Standardized-versus-field-factor comparison is a measurement-validation analysis supporting these questions, not a separate causal claim.

\noindent\textbf{Contributions:}
\begin{enumerate}
\item We make the identification limit explicit: the original $-5.72/-5.24$\,s/week mixed-model estimates are conditional baselines, while crossed-meet penalties and meet-cluster bootstrap show substantial uncertainty once calendar-linked meet difficulty is admitted.
\item We validate measurement without treating the adopted weather rule as an independent discovery. Weak parameter profiles retain Standardized as primary, and a prospective 2025 repeated-venue test shows modest course-factor gains where prior venue data exist.
\item We show that nationals associations are between-program descriptions. Effective Racing Opportunity is algebraically total starts times Hill evenness and is retired as a distinct construct; mechanical maxima, attendance selection, continuous outcomes, and within-team equivalence checks delimit the result.
\end{enumerate}

\noindent\textbf{Coaching takeaway:}
After the opener, race-1-only forecasts do not identify who will improve (2023$\to$2024 SVR $R^2=0.025$ men, $-0.006$ women). Use Standardized times for cross-meet comparison, but do not read a pooled seasonal slope as pure fitness when late-season meets differ systematically. Three- and four-race groups show larger descriptive gains than two-race groups, yet informative continuation and schedule selection prevent a causal prescription. At the program level, larger programs that provide more starts also place higher at nationals; the association is between programs, not evidence that adding a race will improve an athlete or move a team up the standings.

\section{Related Work}
\label{sec:related_work}
Prior work predicts race outcomes from physiology and limited historical results, including matrix-completion and power-law approaches, marathon context models, club XC studies, and Riegel-style distance conversion \cite{jianjun2025predictive,blythe2016prediction,dash2024win,millettpredicting,riegel1981athletic}. Environmental and course factors extend that tradition \cite{cusick2023impact,MaxPerformanceRunning,gutierrez2025real,wilson2024assessing}; LACCTiC-style field factors infer meet difficulty from overlapping athletes rather than weather metadata \citep{mazaheri2026lacctic}. Predictive sports-analytics work also emphasizes data quality, external validity, and coaching interpretability \cite{pietraszewski2025role,rajvsp2025role,rothschild2024predicting,bartosz2025longitudinal,kranzinger2025scoping}.

A second cluster documents gender imbalance in sport science and sex-specific endurance physiology \cite{james2023underrepresentation,anderson2023under,smith2022auditing,besson2022sex,xie2022sex,hunter2023biological}. Scheduling and endurance-workload research links dense competition to recovery, injury risk, and later performance, but collegiate XC race frequency remains understudied \cite{schwarz2024associations,page2023effects,howle2020injury,matos2021variations,lundstrom2025pre}. Appendix~\ref{app:related_work} retains the fuller discussion.

\section{Method}

\subsection{Data}
\label{sec:data}
\begin{figure}[htbp]
\centering
\includegraphics[width=\linewidth]{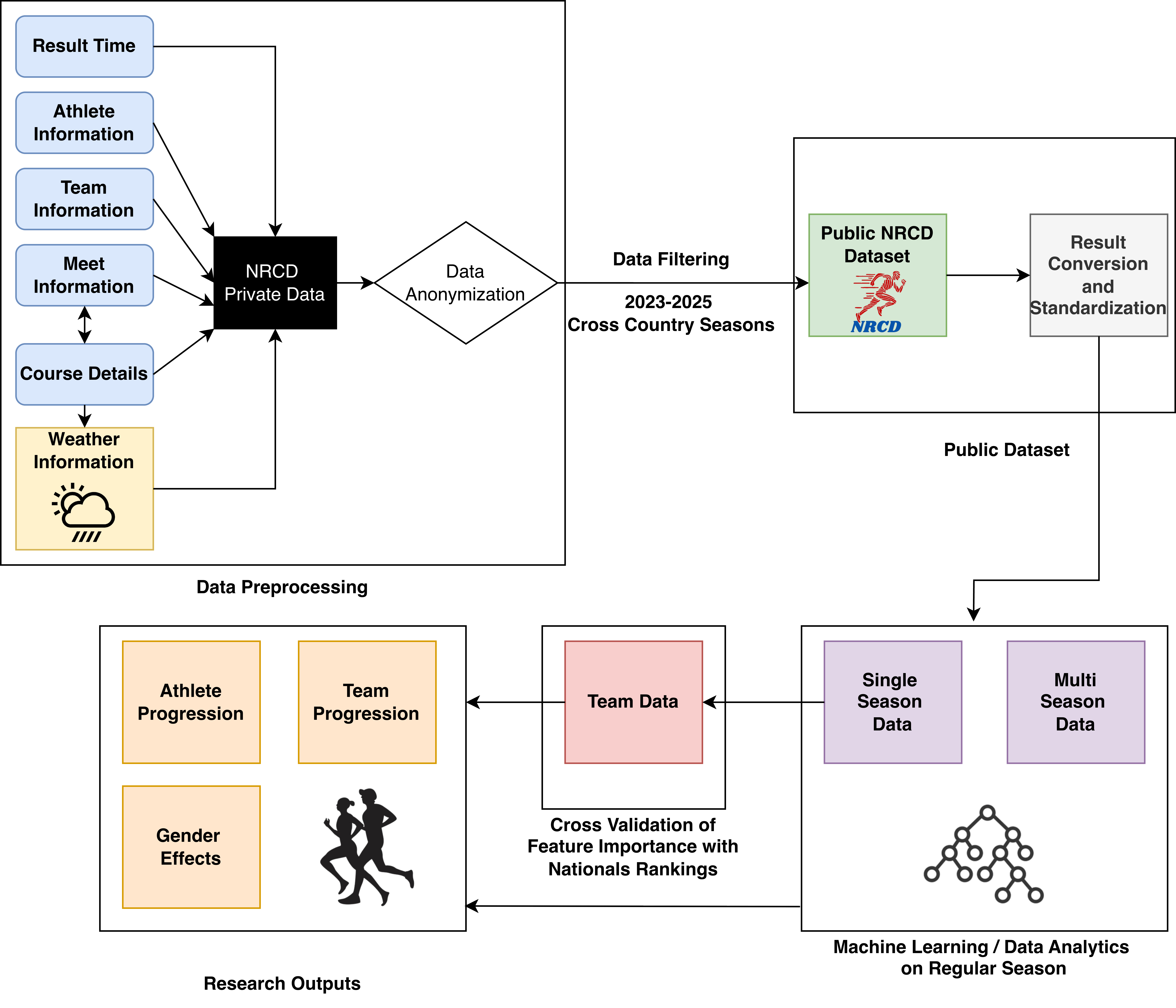}
\caption{NRCD analysis pipeline used in this paper. Private result, athlete, team, meet, course, and weather fields are anonymized into the public NRCD dataset, then Converted Only / Standardized finish times feed athlete and team progression, gender-effect, team-structure, and single- vs.\ multi-season analyses.}
\label{fig:nrcd_pipeline}
\end{figure}

We analyze the comprehensive (current) Cross Country subset of the NRCD. Construction, schema, era definitions, de-identification, and the Converted Only / Standardized formulas are documented in the resource paper \citep{karr2026nrcd}; the public export used here is the Zenodo Full Data Release (v2.1.0) \citep{nrcddataset2026}. Figure~\ref{fig:nrcd_pipeline} summarizes how private NRCD fields become the public export and feed the athlete, team, gender, and single- vs.\ multi-season analyses in this paper.

\noindent\textbf{Inclusion criteria:}
We retain rows whose sport is Cross Country and whose meet \texttt{start\_date} is on or after 2023-08-01 (the comprehensive era). Track and road races are excluded. Within-season analyses additionally drop national-championship meets, since not all athletes can qualify. The hierarchical trajectory uses all $5{,}612$ athlete--seasons with at least two usable Standardized races; the two-point improvement and forecasting analyses require those races to span $\ge 7$ days ($n=5{,}578$). Of $9{,}861$ regular-season athlete--seasons with a Standardized time, $4{,}249$ ($43.1\%$) have a single recorded race and enter neither analysis (Appendix Table~\ref{tab:analytic_sample}). Table~\ref{tab:2023_gender_race_distribution} is a year$\times$gender race-count census on the same nationals-excluded export and can differ slightly in bin totals when a season lacks a usable Standardized time. One-race athletes are slower, later, on smaller rosters, and about half as likely to return next year (Appendix~\ref{app:analytic_sample}), not a random slice of club runners. This paper does not pool the historical NRCD era or NCAA/NAIA/NJCAA varsity exports; the analysis population is comprehensive-era club XC as currently recorded (Appendix~\ref{app:analysis_population}).

We use the comprehensive era rather than the historical one because only the former supports environmental standardization. In the comprehensive XC sample, 99.9\% of results match a \texttt{course\_details} row and 99.9\% have non-null weather conditions, whereas historical XC weather coverage is 27.2\% (Table~\ref{tab:sample_coverage}). The comprehensive era is also comparable in size to prior XC-only releases ($\approx$23k results). Year$\times$gender counts are in Table~\ref{tab:sample_year_gender}.

\begin{table}[htbp]
\centering
\begin{adjustbox}{max width=\linewidth}
\begin{tabular}{lrrrrr}
\toprule
Era & Results & Athletes & Meets & Course details (\%) & Weather (\%) \\
\midrule
Historical XC & 52{,}129 & 17{,}865 & 266 & 37.9 & 27.2 \\
Comprehensive XC & 23{,}360 & 7{,}056 & 280 & 99.9 & 99.9 \\
\bottomrule
\end{tabular}
\end{adjustbox}
\caption{Cross country sample size and metadata coverage by NRCD era. Coverage is the share of results joinable to \texttt{course\_details} (meet + event) with non-null weather when indicated.}
\label{tab:sample_coverage}
\end{table}

\subsection{Performance Standardization}
\label{sec:data_standardization}
Raw finishing times are not comparable across courses. Dataset construction and the Converted Only / Standardized formulas are defined in the NRCD resource paper \citep{karr2026nrcd} and implemented by \texttt{nrcd} \citep{nrcdpackage}; Appendix~\ref{app:std_formulas} writes the equations used here. Converted Only is Riegel distance conversion to $8{,}000$\,m / $6{,}000$\,m. Standardized applies weather, grade, and altitude corrections, each neutral when its inputs are missing (Equation~\ref{eq:std_pieces}). Weather coverage is $\approx 99.9\%$, but gain and loss are present for only about $37\%$ overall and grade availability falls from $11.5\%$ to $3.8\%$ for men and $7.6\%$ to $2.1\%$ for women from early to late weeks. \textbf{Primary analyses use Standardized times}, conditional on the identifying assumption that they remove residual meet difficulty correlated with calendar week.

\noindent\textbf{Sign convention (negative $=$ faster).}
Signed changes and slopes on finish time always use later minus earlier (or time on calendar week), so \textbf{negative $=$ faster}. Hierarchical week slopes, the two-point rate $(t_L-t_1)/\Delta_{\mathrm{days}}$, MixedLM coefficients, and first-to-last totals $\delta=t_L-t_1$ share that finish-time orientation. Other running papers sometimes report ``seconds gained'' as a positive improvement score; we do not mix the two conventions. Unsigned prose such as ``$X$\,s faster'' or ``overstates by $X$\,s'' is a magnitude only.

Writing $a_{ir}:=t^{\mathrm{conv}}_{ir}-t^{\mathrm{std}}_{ir}$ for the per-race environment residual and $\delta=t_L-t_1$ for first-to-last change (negative $=$ faster),
\begin{equation}
\delta^{\mathrm{conv}}-\delta^{\mathrm{std}}
=(t_L^{c}-t_1^{c})-(t_L^{s}-t_1^{s})
=a_L-a_1.
\label{eq:weather_path}
\end{equation}
by definition. When early meets are warmer, $E[a_1-a_L]>0$ and Converted Only is more negative than Standardized (appears to improve more); that gap $E[a_1-a_L]$ is the $15$--$21$\,s inflation in Section~\ref{sec:weather_inflation}. Decomposing $a$ into weather-only vs.\ grade/altitude pieces shows that $\approx 92$--$93\%$ of the seasonal drift is weather. The resource paper's weather coefficient is taken as given; we do not claim a date-conditional re-identification here.

\subsection{Relative finish and field-relative course factors}
\label{sec:field_relative}
Metadata standardization is not the only way to compare heterogeneous XC meets. LACCTiC \citep{mazaheri2026lacctic} adjusts collegiate times by estimating a multiplicative course factor from how the same athletes run across venues. We implement a transparent \emph{club} analogue (not a bit-for-bit reimplementation; no track-PR calibration) and ask whether it should replace Standardized for our improvement estimand.

\noindent\textbf{Relative finish:}
Within each meet$\times$gender field we rank athletes by finish time and map place to a finish percentile in $[0,1]$ (1 $=$ won the field). Athlete--seasons require field size $\ge 5$ at both the first and last race ($n=1{,}971$ women, $3{,}484$ men). We report first-to-last change in Raw, Converted Only, and Standardized times alongside change in finish percentile.

\noindent\textbf{Field factor $\alpha$:}
Gender-separated damped median EM (Equation~\ref{eq:field_em}; Appendix~\ref{app:field_em}) estimates athlete fitness $f_i$ and meet factor $\alpha_j$ on a chosen time $t^{(c)}\in\{t^{\mathrm{raw}},t^{\mathrm{conv}},t^{\mathrm{std}}\}$, with $\operatorname{median}_j\alpha_j=1$. Adjusted time is $\tilde t_{ij}=\alpha_j t^{(c)}_{ij}$ (smaller $\alpha$ on slower meets). We compare season $\delta=t_L-t_1$ under field-adjusted vs.\ Standardized times and correlate meet hardness $h_j=1/\alpha_j$ with the NRCD environment residual $a=t^{\mathrm{conv}}-t^{\mathrm{std}}$ and with observed temperature. The EM never takes weather or elevation as covariates; those enter only through $t^{(c)}$.

\noindent\textbf{Finish-time pairing (isolation check):}
Fitness is held fixed in the EM window, so fitness gains can load into $\alpha_j$. We report leftover $\rho(1/\alpha,T)$ (partial given day of year) after Converted Only vs.\ Standardized, plus a joint\_tdoy sensitivity (Appendix~\ref{app:field_em}).

\noindent\textbf{Head-to-head criteria:}
We score methods on next-raw MAE, within-meet place Spearman, held-out residual SD, split-half ability reliability, and leftover $\rho(1/\alpha,T)$ (partial given day of year). The \textbf{Std-then-$\alpha$} hybrid (Equations~\ref{eq:std_pieces}--\ref{eq:prerace}) Standardizes available metadata and then fits field $\alpha$ on that time. Because the target-meet $\alpha$ and raw-to-Standardized scale use results from the target meet, its time and place scores are retrospective reconstruction benchmarks, not deployable pre-race forecasts. Standardized remains primary for within-season progression. Full EM and pairing details are in Appendix~\ref{app:field_em}.

\subsection{Individual improvement models}
\label{sec:ml_estimand}
\noindent\textbf{Primary individual model (hierarchical trajectory):}
\label{sec:hier_trajectory}
Rather than summarize a season by the two-point rate $(t_L-t_1)/\Delta$, we model every nationals-excluded Standardized race for athlete--season $is$ as
\begin{equation}
\begin{aligned}
t_{isj}
&=
\beta_0 + \beta_1\,\mathrm{week}_{isj}
+ \text{year FE}
+ u_{0,is} + u_{1,is}\,\mathrm{week}_{isj}
+ \varepsilon_{isj},
\\
\begin{pmatrix}u_{0,is}\\ u_{1,is}\end{pmatrix}
&\sim
\mathcal{N}\!\bigl(0,\Sigma_u\bigr),
\qquad
\varepsilon_{isj}\sim\mathcal{N}(0,\sigma^2),
\end{aligned}
\label{eq:hier_trajectory}
\end{equation}
with $\mathrm{week}_{isj}$ equal to days from the gender$\times$year mean race date, divided by 7 (so $0$ is mid-season for that sex and year, negative is earlier, positive is later), and athlete--season random intercept and random slope $(u_{0,is},u_{1,is})$. Centering puts the origin at the typical meet date rather than January~1 or season opener; it does not mean ``week zero of the calendar.'' Negative $\beta_1$ means the field gets faster as the season progresses past mid-season. The primary pooled fit allows year intercepts but a common week slope; we also report per-year slopes and a week$\times$year interaction because slopes can differ across seasons. Using every race reduces the reliability problem of the two-point rate, shrinks noisy athletes toward the group mean slope, and absorbs year-level intercepts in the fixed effects. Gender-separate fits are primary.

\noindent\textbf{Coach-facing forecast after meet 1:}
\label{sec:early_forecast_method}
The primary forecast uses only opening Standardized time, trains on 2023, and tests on 2024. It excludes full-season race count, schedule span, and all future race timing, so it has no schedule leakage. Ridge and RBF SVR predict the eventual first-to-last rate defined below. After-meet-2 models are secondary early-window sensitivities; the six-family, full-season suite is retained only as appendix provenance.

\noindent\textbf{Supporting estimand (two-point rate):}
For athlete $i$ in season $s$ with ordered regular-season races (nationals excluded), let $t_{i,s,1}$ and $t_{i,s,L}$ be the first and last Standardized times and $\Delta_{i,s}$ the days between those meets (not the first-to-last time change). The two-point improvement rate is
\begin{equation}
y_{i,s} = \frac{t_{i,s,L} - t_{i,s,1}}{\Delta_{i,s}}
\label{eq:improvement_rate}
\end{equation}
(seconds per day; negative $=$ faster, matching $\delta=t_L-t_1$ and the hierarchical week slope). We treat $y_{i,s}$ as a coach-facing summary and forecast target, not as the primary within-season estimand. Retrospective full-season models exclude the final time but can use schedule descriptors and therefore are provenance analyses, not the primary coach forecast.

\noindent\textbf{Relation to the resource paper's illustrative $R^2$:}
The resource paper \citep{karr2026nrcd} reported illustrative RF/GB $R^2$ of roughly $0.5$--$0.74$ as a paired Converted vs.\ Standardized check, not as out-of-year forecasting. Absolute $R^2$ there is high by construction (features include last-race summaries of the label). With the outcome race excluded, held-out $R^2$ here falls to $\approx 0$ (Table~\ref{tab:gender_primary_performance}); the two results are complementary (Appendix~\ref{app:ml_spec}).

\noindent\textbf{Descriptive dose--response:}
Table~\ref{tab:dose_response} reports first-to-last Standardized change in total seconds by race-count bin. Two robustness checks (the pre-final trajectory slope and first-to-fastest improvement, Figure~\ref{fig:combined_improvement}) agree qualitatively that, among slower starters, gains look larger when the binning race is an endpoint (Galton/regression to the mean \cite{barnett2005regression}); they can differ in sign for near-flat athletes (Appendix~\ref{app:ml_spec}).

Appendix~\ref{app:ml_spec} preserves the six-model full-season provenance (OLS, Ridge, Lasso, Random Forest \cite{breiman2001random}, Gradient Boosting \cite{friedman2001greedy}, and SVR \cite{awad2015support}), its temporal metrics, and feature audit. Those retrospective results do not displace the race-1-only forecast.

\noindent\textbf{Temporal validation and signal extraction:}
Primary: train on 2023, test on 2024. Because 2023 is the first comprehensive-era season and may be sparsely covered, we also report a 2024$\to$2025 sensitivity for the compact SVR, early-window check, and hierarchical last-race transfer. Secondary (RQ2): train 2023$\to$test 2025; train 2023+2024$\to$test 2025 \cite{roberts2017cross}. Following the resource paper's use of gender-specific target distances and because secondary associations can differ by gender, \textbf{primary models are trained separately for men and women on Standardized times}. Pooled models are retained only as diagnostics; Converted Only and unadjusted-time fits are sensitivity analyses.

Held-out $R^2$ alone is hard to interpret when the outcome is noisy. If the improvement slope has split-half reliability $\rho_{yy'}$, classical attenuation implies $R^2\le\rho_{yy'}$. We therefore also report the Signal Extraction Ratio
\begin{equation}
\mathrm{SER}:=\frac{R^2_{\mathrm{heldout}}}{\rho_{yy'}},
\label{eq:ser}
\end{equation}
the fraction of \emph{reliable} outcome variance captured (negative $R^2$ implies non-positive SER). Reliability estimation and related noise-floor checks are in Appendix~\ref{app:null_diagnostics}.

\noindent\textbf{Feature exclusion policy:}
\label{sec:feature_exclusion}
Because $y=(t_L-t_1)/\Delta$ is built from race times, we exclude endpoint and derived leakage (\texttt{last\_time}, \texttt{total\_improvement} and aliases), exact duplicate columns, and within-sex gender encodings, then prune near-duplicate retained covariates ($|\mathrm{corr}|\ge0.95$) to 17 within-sex predictors. Add-back of \texttt{last\_time} raises men's $R^2$ by $+0.44$; compact vs.\ legacy-full $\Delta R^2=-0.007$ (Appendix~\ref{app:ml_spec}, Table~\ref{tab:feature_exclusion}).

\noindent\textbf{Team-level analysis:}
For each team--season we compute maximum race count among roster athletes of that gender (a meet counts only if $\ge 3$ same-gender teammates competed) and roster depth ($\#$ athletes with $\ge 3$ starts). To summarize depth and intensity in one scalar, let $\{n_j\}_{j=1}^{N}$ be start counts on the roster ($N\ge 1$, $n_j\ge 0$, $\sum_k n_k>0$), $p_j=n_j/\sum_k n_k$, Shannon entropy $H=-\sum_{j:p_j>0}p_j\log p_j$ (natural log; $0\le H\le\log N$), $\bar n=N^{-1}\sum_j n_j$, and
\begin{equation}
\mathrm{ERO}
:=
e^{H}\,\bar n
=
\left(\sum_{j=1}^{N}n_j\right)
\left(\frac{e^H}{N}\right)
=
\text{total starts}\times\text{Hill evenness}.
\label{eq:ero}
\end{equation}
Here $e^{H}$ is the effective number of equally racing athletes and $e^H/N$ is Hill evenness. Because Equation~\ref{eq:ero} is an exact reparameterization of total starts and evenness, we do not treat ERO as a distinct construct. We retain its held-out discrimination descriptively, test nationals outcomes and attendance selection, and separate between- from within-program effects with a Mundlak decomposition (Appendix~\ref{app:team_selection}).

\noindent\textbf{Multi-season and gender analyses:}
For RQ2, athletes are retained across adjacent seasons when $|\Delta$ races$|\le 1$; we evaluate the pre-specified SVR temporally and summarize changes in best Standardized time. For RQ3, we compare athlete counts and starts per recorded competitor by sex, then fit one pooled random-intercept model with sex$\times$race-count interactions. Full specifications and sensitivity tables remain in the appendix.

\section{Results}

\subsection{Runners' Improvement Over a Season}
RQ1 asks whether runners improve within a season on Standardized times (Section~\ref{sec:data_standardization}). The primary answer uses the hierarchical trajectory in Equation~\ref{eq:hier_trajectory}; the two-point rate (Equation~\ref{eq:improvement_rate}) and its leakage-controlled SVR forecast are supporting. Unless noted, reported $R^2$ values use Standardized times under temporal validation.

\subsubsection{Hierarchical trajectory (primary)}
\label{sec:hier_results}
Gender-separate MixedLMs of every Standardized race give slopes of $-5.72$\,s/week for men (95\% CI $[-6.85,-4.59]$; $n=3{,}563$ athlete--seasons, $9{,}907$ races) and $-5.24$\,s/week for women ($[-6.45,-4.03]$; $n=2{,}049$, $5{,}590$ races). These are \emph{conditional baseline estimates}, not unqualified fitness effects: they identify improvement only if Standardized times remove residual meet difficulty associated with week.

That assumption is consequential. Late weeks contain larger fields (early/late means $95.6/152.4$ men; $62.7/93.3$ women), championship-like meets rise from about $2\%$ to $55\%$, and grade availability declines. In sparse crossed athlete--season/meet models, the week coefficient ranges from $+1.34$ to $-5.48$\,s/week for men and $-3.97$ to $-5.84$ for women as the meet-effect penalty increases from $0.1$ to $100$. At the reporting anchor (penalty $10$), estimates are $-3.95$ and $-4.93$; excluding all championship-like meets gives $-3.47$ and $-2.91$. Including nationals gives $-3.73$ and $-5.80$, or $-3.41$ and $-4.53$ with a nationals indicator. A meet-cluster bootstrap is still less precise: $-1.68$\,s/week $[-5.35,1.97]$ for men and $-1.31$ $[-4.47,1.94]$ for women. The sign is therefore not robustly separated from zero under meet-level resampling, especially for men.

The fitted latent random-slope SD is $12.0$\,s/week for men and $6.0$ for women, but a large latent SD does not imply well-measured individual slopes. About $47\%/50\%$ of athlete--seasons have exactly two races. Approximate empirical-Bayes slope reliability averages only $0.47$ for two-race men and $0.11$ for two-race women, rising with race count. The model can estimate a broad latent distribution while any given athlete's slope remains uncertain because races are few, closely spaced, and noisy.

Those pooled slopes average over heterogeneous years (Table~\ref{tab:hier_per_year}). Men's 2024 and 2025 slopes are clearly negative ($-7.1$ and $-11.4$\,s/week); the 2023 slope is positive ($+3.1$), indicating slower times later in that season. Women's point estimates are negative in all three seasons (2023's CI includes zero; 2024 is steepest). A thinner 2023 meet calendar (no August meets; fewer late-season meets) is the primary caution for that year; Converted Only 2023 men are near flat while Standardized can tip the sign (Appendix~\ref{app:supp_results}).

Define the variance ratio as the variance of the two-point rate divided by the variance of the hierarchical BLUP total slope, with both in seconds/day. It is about $54$ for men and $403$ for women; because BLUPs are shrunken, this is descriptive rather than a reliability coefficient. Spearman agreement is only $\rho\approx0.32$--$0.37$. A quadratic week term is near zero for men. For women it is negative ($-0.93$\,s/week$^2$): both early and late fitted values differ symmetrically from the linear path, with a relatively flat early segment and faster late decline that may reflect taper or championship composition rather than physiology alone.

\begin{table}[htbp]
\centering
\begin{adjustbox}{max width=\linewidth}
\begin{tabular}{llrrr}
\toprule
Gender & Year & $\beta_{\mathrm{week}}$ (s/week) & 95\% CI & $n$ athlete--seasons \\
\midrule
Men & 2023 & $+3.10$ & $[1.26,\,4.95]$ & 1151 \\
Men & 2024 & $-7.06$ & $[-8.50,\,-5.63]$ & 1170 \\
Men & 2025 & $-11.38$ & $[-13.92,\,-8.83]$ & 1242 \\
\midrule
Women & 2023 & $-1.81$ & $[-3.91,\,0.28]$ & 632 \\
Women & 2024 & $-11.03$ & $[-13.13,\,-8.93]$ & 671 \\
Women & 2025 & $-3.30$ & $[-4.91,\,-1.69]$ & 746 \\
\bottomrule
\end{tabular}
\end{adjustbox}
\caption{Per-year hierarchical week slopes (Standardized time; athlete--season random intercept and slope; no pooled year FE). Negative $=$ getting faster. Men's improvement is concentrated in 2024--2025; the positive 2023 estimate indicates slower times later in that season. Women's point estimates are negative every year (2023 CI includes zero).}
\label{tab:hier_per_year}
\end{table}

\begin{figure}[htbp]
\centering
\includegraphics[width=0.95\linewidth]{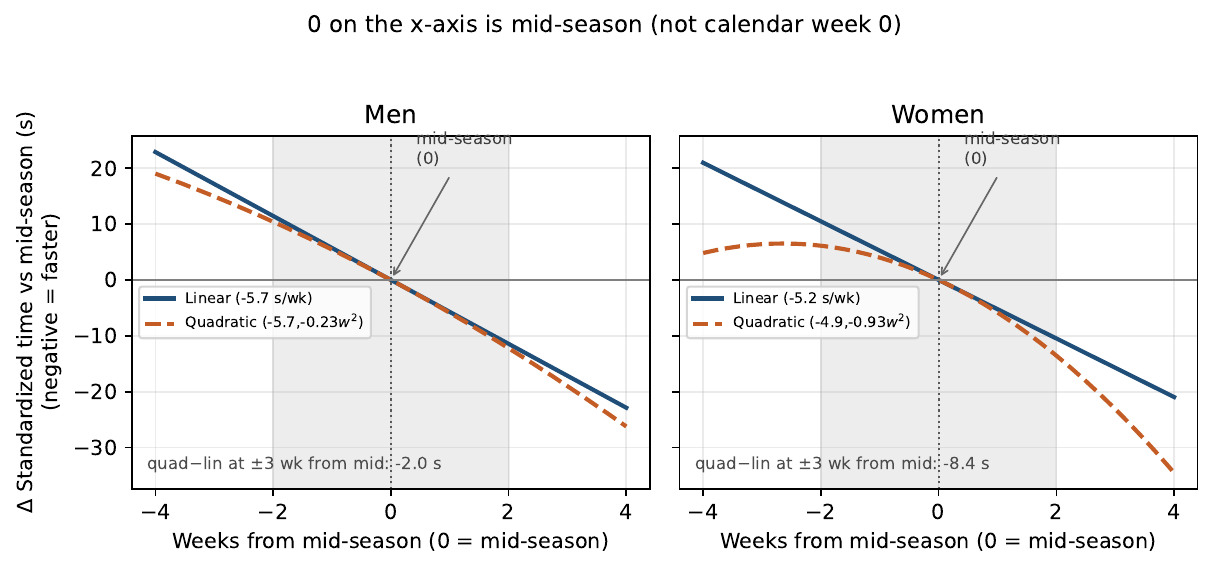}
\caption{Fitted population trajectories. Horizontal axis: weeks from mid-season; the dotted line and label mark \textbf{0 $=$ mid-season} (gender$\times$year mean race date), not calendar week~0 (negative $=$ earlier meets, positive $=$ later). Curves are demeaned at mid-season so the plot shows change relative to that midpoint, not absolute finish time. Shaded band marks a typical $\pm 2$\,week span around mid-season. Men's linear and quadratic paths nearly coincide. For women, the negative quadratic term places both early and late fitted times about $8$\,s below the linear path at $\pm3$ weeks; the resulting shape is relatively flat early and declines more steeply late, a pattern that may reflect taper or championship-meet composition.}
\label{fig:hier_trajectories}
\end{figure}

Opening time and roster size have descriptive week interactions in level-2 models (Appendix~\ref{app:trajectory_identification}): a 60\,s slower opener associates with $-3.79$\,s/week for men and $-2.37$ for women; ten additional roster athletes associate with $-1.51$ for men and $-0.23$ for women. These are not explanations of latent heterogeneity. Adding the interactions increases, rather than reduces, fitted random-slope variance (men $145\to373$; women $37\to185$\,s$^2$/week$^2$), so optimizer/variance reallocation and residual structure preclude variance-explained claims.

Continuation is informative: one-race seasons are selected out, and an inverse-continuation-weight sensitivity shifts meet-cluster slopes from $-1.68$ to $-2.08$\,s/week for men and from $-1.31$ to $-1.18$ for women. The model cannot recover unrecorded races or remove unmeasured continuation factors. Span-adjusted associations also remain descriptive: each additional observed race associates with $-18.4$\,s total change for men and $-9.9$ for women after season span adjustment; race density per 30 observed days associates with $-1.13$\,s/day for men and $-0.14$ for women (women's interval includes zero). These are selected-dose associations, not effects of assigning another race.

\subsubsection{Coach-facing forecast after meet 1}
\label{sec:feature}
\label{sec:early_window}
The primary coach forecast uses opening Standardized time only and no full-season schedule information. Training on 2023 and testing on 2024, SVR gives $R^2=0.025$ for men ($n=1{,}164$) and $-0.006$ for women ($n=670$), with MAE $3.66$ and $4.16$\,s/day. This is the information actually available after meet~1 and does not leak race count, season span, or future timing. The 2024$\to$2025 sensitivity remains weak for SVR ($0.094$ men, $0.002$ women), although men's Ridge reaches $0.267$, underscoring split and learner instability. After-meet-2 results and the six-learner full-season suite are appendix provenance, not the primary forecast.

\begin{figure}[htbp]
\centering
\includegraphics[width=0.95\linewidth]{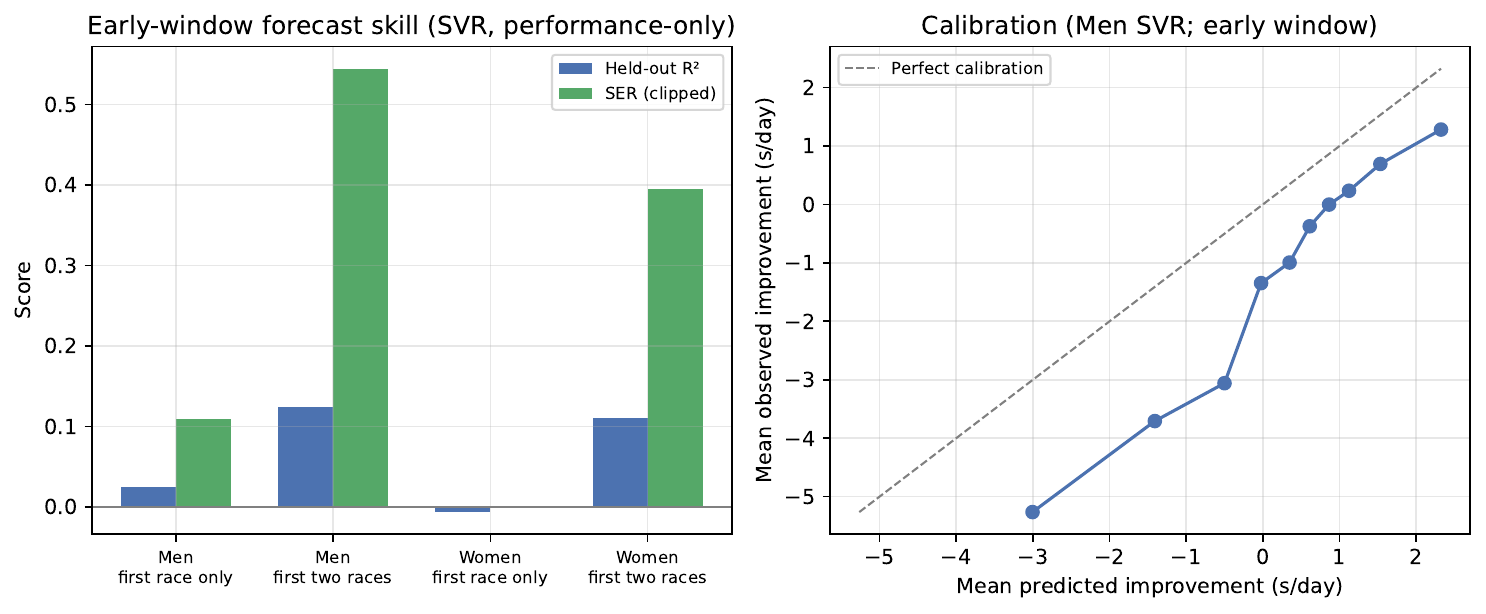}
\caption{Early-window forecast skill (train 2023$\to$test 2024). The meet-1 specification uses opening Standardized time only and excludes full-season schedule information. Skill is near noise; meet-2 estimates are secondary sensitivities on the selected $\ge3$-race subgroup.}
\label{fig:early_window_decision}
\end{figure}

RF importances are not interpreted as drivers. Add-back of \texttt{last\_time} raises men's $R^2$ by $+0.44$ (leakage; Section~\ref{sec:feature_exclusion}). Appendix~\ref{app:null_diagnostics} gives the signal ceiling: slope split-half reliability only $0.23$--$0.28$, $\mathrm{SER}=0.19$ (men; non-positive women), permutation $p=0.001$ with no useful MAE skill, and tail-classification AUC that collapses once baseline-coupled features are removed. Converted Only / unadjusted SVR sensitivities are likewise near zero ($R^2\approx 0.04$), so Standardized remains primary for comparability, not for maximizing $R^2$.

\FloatBarrier

\subsubsection{Exploratory schedule and continuation}

\noindent\textbf{Race-count dose--response (exploratory):}
\label{sec:dose_response}
Median Standardized first-to-last change deepens modestly from two to three--four races (men $-11\!\to\!-18\!\to\!-20$\,s; women $-14\!\to\!-20\!\to\!-29$\,s), then flattens in the sparse $5+$ bin (Appendix Table~\ref{tab:dose_response}; cell $n$ ranges from 136--1{,}642 for men and 72--1{,}014 for women). Quartile strata concentrate among slower starters and are affected by regression to the mean when binning on $t_1$; a valid race-2$\to$last check leaves only a modest Q4--Q1 gap ($d\approx 0.29$--$0.31$; Appendix Figure~\ref{fig:rtm_vs_ability}). These associations are exploratory, not causal schedule prescriptions. A mixed-model quadratic on race count shows some diminishing returns for men near the $5+$ bin (Appendix~\ref{app:rq1_mixed}).

\subsection{Measurement validation}

\noindent\textbf{Weather-rule magnitude and profile:}
\label{sec:weather_inflation}
The adopted $H=T+D$ rule with $k=0.0016$ and threshold $H_0=100$ implies that Converted Only shows $21.2$\,s more first-to-last gain for men (95\% CI $[20.4,22.0]$) and $15.4$\,s for women ($[14.5,16.3]$) than Standardized. These are magnitudes implied by applying the adopted adjustment, not an independent discovery of the weather effect. Deterministic meet-block profiles prefer $k=0.0014$ and $H_0=94$, but the one-SE sets are broad ($k\in[0,0.004]$, $H_0\in[80,130]$), and the profiles vary one parameter while holding the other fixed. Standardized therefore remains primary, without claiming joint parameter identification.

\noindent\textbf{Prospective repeated-venue course factors:}
\label{sec:field_relative_results}
Course factors fit only on 2023--2024 are evaluated on 2025. For races at seen courses, MAE falls from $77.1$ to $73.7$\,s for men (1{,}893 predictions at 51 meets) and from $62.5$ to $58.3$ for women (978 predictions at 38 meets). Seen-course coverage is $84.7\%$ of men's and $76.4\%$ of women's next-race predictions (82.5\% and 75.2\% of all 2025 races). Unseen courses receive $\alpha=1$, exactly matching the Standardized-only course assumption. This prospective repeated-venue result is the primary course-factor validation.

\noindent\textbf{Retrospective reconstruction:}
Full-sample field $\alpha$ remains useful only as a retrospective reconstruction. It improves next-race times and order but uses target-meet results, and therefore is neither a deployable pre-race forecast nor a progression scale. Standardized times show within-season change while mean finish percentile is near zero, as expected for a zero-sum field measure. Full comparisons remain in Table~\ref{tab:course_method_horse} and Appendix~\ref{app:field_em}.

\begin{table}[htbp]
\centering
\begin{adjustbox}{max width=\linewidth}
\begin{tabular}{llccc}
\toprule
Criterion & Prefer & Men & Women & Notes \\
\midrule
Next raw-time MAE (s) & Field $\alpha$ / hybrid & $50.1$ / $49.6$ vs.\ $71.6$ & $45.7$ / $45.3$ vs.\ $66.0$ & vs.\ Standardized; hybrid $=$ Std-then-$\alpha$ \\
Within-meet place Spearman & Field $\alpha$ / hybrid & $0.899$ / $0.901$ vs.\ $0.889$ & $0.909$ / $0.909$ vs.\ $0.902$ & matched last-race fields \\
Split-half ability $\rho$ & Field $\alpha$ & $0.94$ vs.\ $0.90$ & $0.95$ vs.\ $0.91$ & raw finish-time reliability \\
Leftover $\rho(1/\alpha,T)$ & Standardized & $-0.05$ vs.\ $0.31$ & $-0.01$ vs.\ $0.31$ & vs.\ Converted-then-$\alpha$ \\
Partial $\rho(1/\alpha,T\mid\mathrm{doy})$ & Standardized & $-0.05$ vs.\ $0.27$ & $-0.03$ vs.\ $0.25$ & date-orthogonal \\
$\rho(1/\alpha,\mathrm{doy})$ & Standardized & $0.01$ vs.\ $-0.17$ & $-0.04$ vs.\ $-0.22$ & calendar track \\
Season $\delta$ rank agreement & -- & $\rho=0.77$ & $\rho=0.70$ & field-adj vs.\ Std \\
\bottomrule
\end{tabular}
\end{adjustbox}
\caption{Retrospective metadata Standardized vs.\ field-$\alpha$ reconstruction on club XC. The circular environment-residual-versus-temperature comparison is omitted. Target-meet factors use target-meet results, so these rows are not deployable pre-race forecasts; the prospectively trained 2025 result is primary.}
\label{tab:course_method_horse}
\end{table}

\begin{figure}[htbp]
\centering
\includegraphics[width=0.95\linewidth]{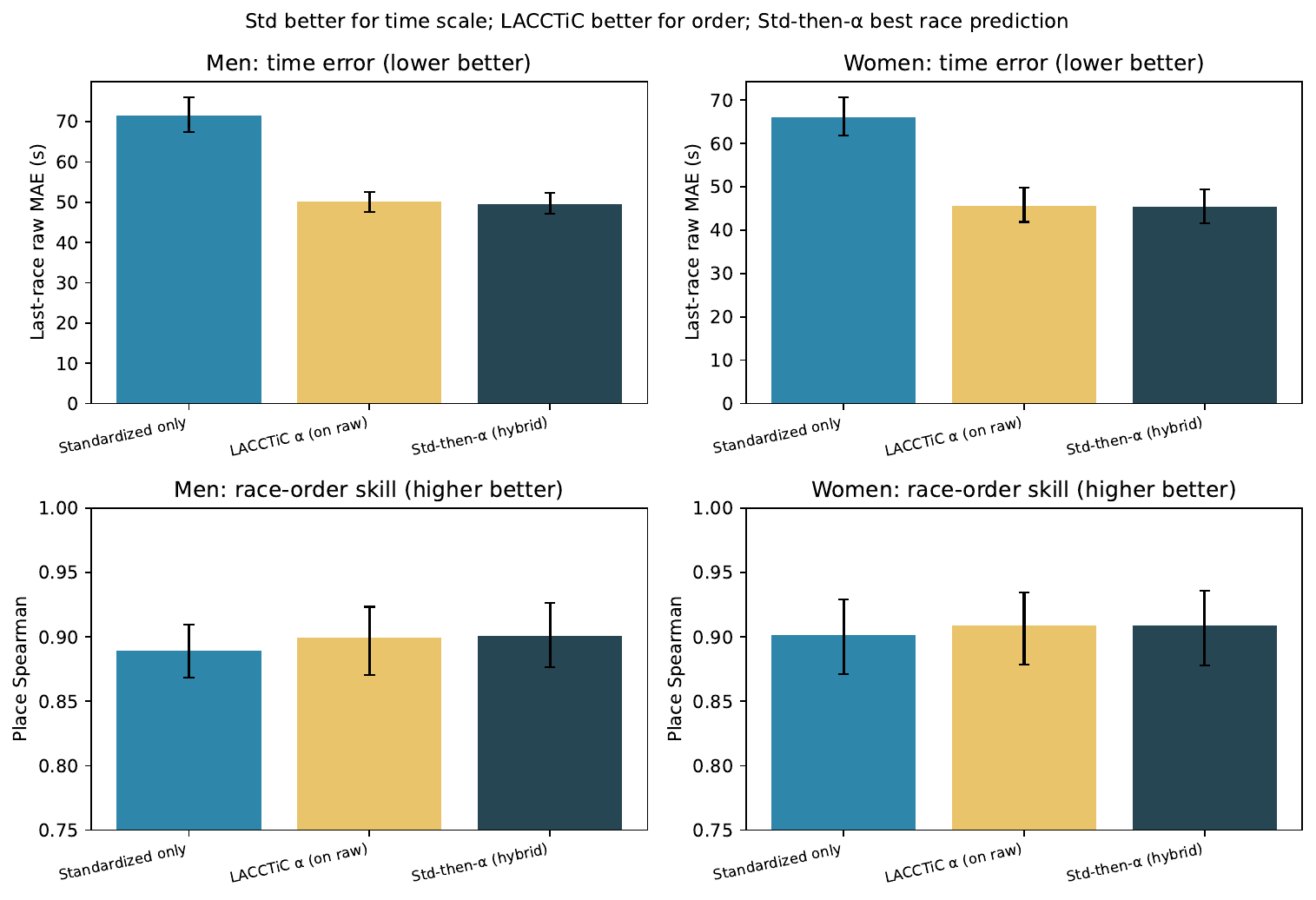}
\caption{Matched last-race sample: Standardized alone, LACCTiC-style $\alpha$ on raw, and Std-then-$\alpha$ hybrid. Top: next-raw MAE (lower better). Bottom: within-meet place Spearman (higher better).}
\label{fig:std_lacctic_race}
\end{figure}

\subsection{Program structure and nationals}
\label{sec:team_results}
Programs that race more place higher in cross-section; racing more \emph{causing} success is not supported (Figure~\ref{fig:team_dag}). The continuous top-15 OR is $2.56$ per SD max races, while top-15 is a secondary dichotomization. Among 149 ranked top-25 rows, each SD of max race count associates with 1.63 places better rank ($95\%$ CI $0.26$ to $3.01$ better). Among 257 team--season--gender rows with at least five valid nationals times, it associates with $0.27$ SD faster top-five performance ($[0.12,0.42]$). No official team-score field exists, so no team scores are constructed or imputed.

\noindent\textbf{Selection robustness:}
\label{sec:team_selection}
A Mundlak within--between LPM puts $+9.1$\,pp per max race between programs and $-0.2$\,pp within. Nationals attendance itself is selected: 334 panel rows have any nationals athlete and 257 have at least five, so continuous outcomes are conditional on reaching the meet. Within-team first differences cover all 328 consecutive qualifying-team pairs; the $\Delta$max-race slope is $-0.87$\,pp per race ($[-4.22,2.49]$), and TOST at $\pm5$\,pp gives $p=0.008$, supporting practical equivalence at that margin. Adding roster size attenuates max-race ORs and collapses depth.

\noindent\textbf{Mechanics and ERO:}
Even under equal random allocation of starts while fixing roster and total volume, the expected maximum correlates $r=0.914$ with the observed maximum. Thus max race count partly reflects arithmetic opportunity. Likewise, ERO is exactly total starts times Hill evenness (Equation~\ref{eq:ero}); once total starts are conditioned on, evenness adds little (OR $1.14$ $[0.79,1.66]$), and conditional on roster size plus mean intensity its OR is $0.98$ $[0.65,1.49]$. We therefore retire ERO as a distinct construct. Its held-out 2025 AUC of $0.836$ remains a descriptive prediction summary, not evidence for an additional mechanism.

\begin{figure}[htbp]
\centering
\begin{tikzpicture}[
  font=\scriptsize,
  box/.style={draw, rounded corners=2pt, align=center, inner sep=3.5pt, minimum width=2.35cm},
  arr/.style={-{Stealth[length=1.6mm]}, thick}
]
  \node[box, fill=gray!12] (res) at (0,2.15) {Program resources\\(budget, size)};
  \node[box, fill=gray!12] (coach) at (4.15,2.15) {Coaching /\\recruiting};
  \node[box, fill=gray!12] (ath) at (8.3,2.15) {Athlete\\quality};
  \node[box, fill=blue!8] (opp) at (4.15,0.55) {Racing opportunity\\(max races, depth, ERO)};
  \node[box, fill=orange!12] (place) at (4.15,-1.05) {Nationals placement};
  \draw[arr] (res) -- (opp);
  \draw[arr] (coach) -- (opp);
  \draw[arr] (ath) -- (opp);
  \draw[arr] (opp) -- (place);
  \draw[arr, dashed] (res.south) to[out=-90,in=180] (place.west);
  \draw[arr, dashed] (ath.south) to[out=-90,in=0] (place.east);
  \draw[arr, dashed] (coach.east) to[out=0,in=40] (place.east);
\end{tikzpicture}
\caption{Confounding for the team result. Observed racing opportunity sits on paths from resources, coaching, and athlete quality, which also point at placement. Cross-sectional RR/OR estimate the association along all of those paths. Within-team first differences difference out stable program factors and are null in these data (Appendix~\ref{app:team_selection}). The figure is a statement of the identification problem, not a fitted structural model.}
\label{fig:team_dag}
\end{figure}

\begin{table}[htbp]
\centering
\begin{adjustbox}{max width=\linewidth}
\begin{tabular}{lll}
\toprule
Analysis & Estimate (95\% CI) & Note \\
\midrule
Pooled RR, $\ge 3$ / $\ge 4$ / $\ge 5$ races & $1.32$ / $2.09$ / $3.04$ & dose-monotone; $\ge 4$: $[1.42,3.06]$ \\
Continuous OR / SD max races (GEE) & $2.56$ $[1.78,3.67]$ & top-15; $p<10^{-6}$; $n=573$ \\
Pooled RR by top-$k$ at $\ge 4$ ($k=5\to25$) & $5.71\to3.28$ & all CIs exclude 1 \\
Cont.\ OR / SD by top-$k$ & $\approx 2.7$--$3.6$ & stable across cuts \\
Depth OR / SD ($\#$ athletes $\ge 3$ starts) & $1.74$ $[1.26,2.42]$ & conditional on max races; $p=0.0009$ \\
ERO OR / SD; OOS AUC & $2.70$ $[2.12,3.44]$; $0.84$ & roster size / total starts AUC $0.83$/$0.84$ \\
Max-race OR / SD after roster-size control & $1.69$ $[1.05,2.71]$ & GEE; attenuates vs.\ $2.56$ alone \\
Depth OR after size control & $\approx 1.05$ (n.s.) & joint with max races; size dominates \\
Non-top-7 depth OR alone / $+$ size & $2.36$ / $1.11$ (n.s.) & beyond season-best top~7, $\ge 3$ starts \\
Within-team $\Delta$max races $\to$ $\Delta$top15 & $-0.9$\,pp/race $[-4.2,2.5]$ vs CS $7.0$\,pp/race & TOST $\pm 5$\,pp/race (max races); $n=328$ \\
\bottomrule
\end{tabular}
\end{adjustbox}
\caption{Descriptive team associations. Top-15 is secondary to continuous rank and top-five performance; attendance conditions who has those outcomes. ERO is reported for held-out discrimination but is algebraically total starts times Hill evenness, not a distinct construct.}
\label{tab:team_robust}
\end{table}

\begin{figure}[htbp]
\centering
\includegraphics[width=0.95\linewidth]{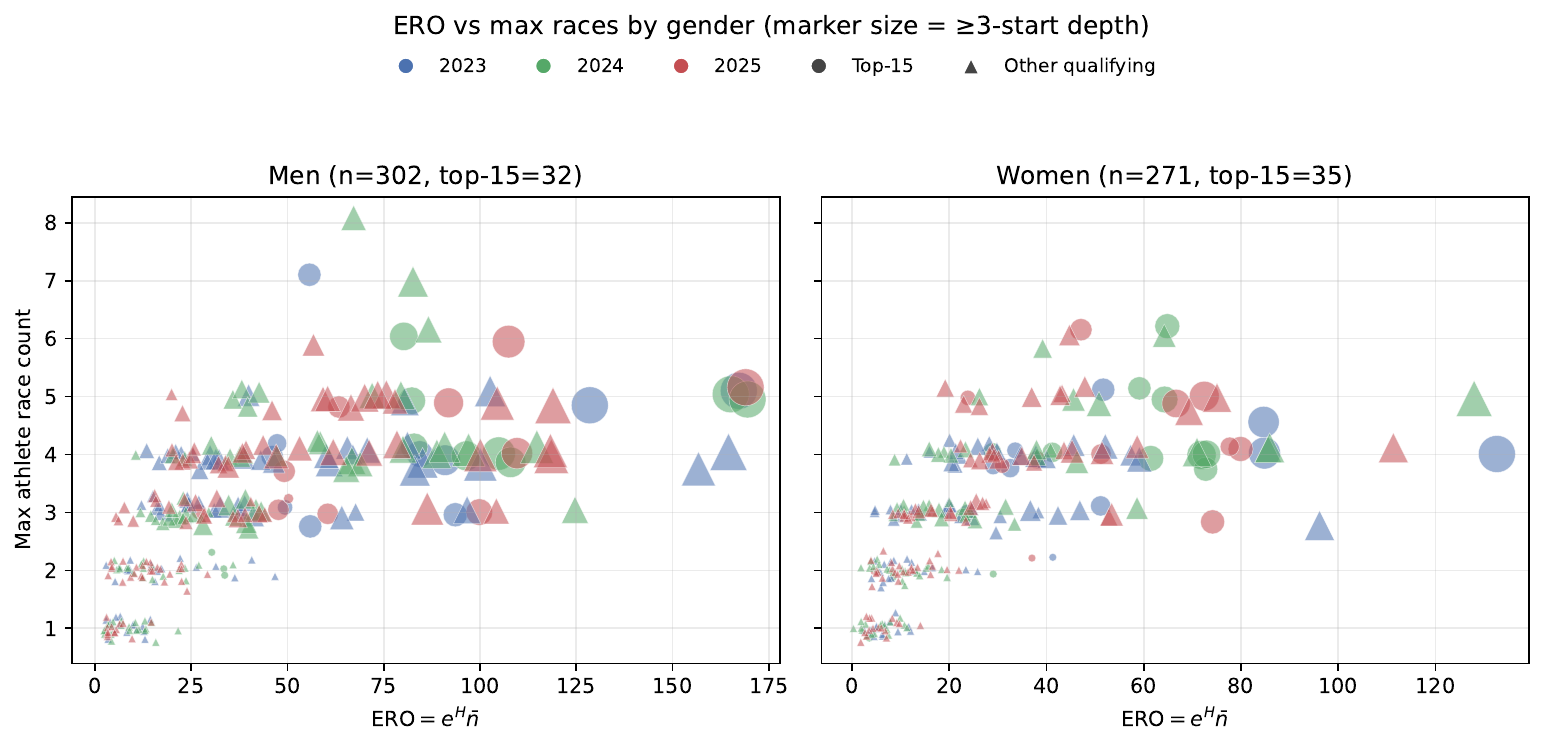}
\caption{Team-season scatter of max race count versus the ERO reparameterization (Equation~\ref{eq:ero}), split by gender. Color $=$ year (2023--2025); circles $=$ nationals top~15, triangles $=$ other qualifying; marker area scales with $\ge 3$-start roster depth. Held-out AUCs are descriptive; ERO is not interpreted as a distinct mechanism.}
\label{fig:ero_vs_maxrace}
\end{figure}

\noindent\textbf{Secondary descriptive results:}
Multi-season forecasting is non-positive on the primary 2023$\to$2024 split (SVR $R^2=-0.09$ men, $-0.08$ women), and every interval for mean adjacent-season best-time change includes zero. Men account for about $60$--$63\%$ of recorded athletes, but starts per recorded competitor are similar by sex. These results describe the filtered panel and observed participation, not collegiate careers or unobserved roster entry. Full forecasting families, retention, and participation tables remain in Appendix Tables~\ref{tab:rq2_splits}--\ref{tab:2023_gender_race_distribution}.

\label{sec:rq3}
The apparent 2023 men's count difference, 1{,}927 in Table~\ref{tab:sample_year_gender} versus 1{,}920 in Appendix Table~\ref{tab:total_unique_results}, is expected: the former is the comprehensive sample summary, while the latter applies the regular-season filters used for the participation appendix. The same distinction explains small differences in other year--sex cells.

\FloatBarrier

\section{Discussion}
\label{sec:discussion}

The hierarchical trajectory gives conditional baseline slopes of $-5.72$ and $-5.24$\,s/week, not unqualified fitness effects. Identification requires Standardized times to remove meet difficulty that co-moves with week. The crossed-meet penalty grid and meet-cluster bootstrap show that this condition is empirically consequential: men's sign changes under weak meet shrinkage, and both bootstrap intervals include zero. Late-season championship concentration, larger fields, and lower grade availability explain why calendar week cannot be interpreted in isolation.

The coaching triage question is answered by the race-1-only experiment, not by retrospective full-season models. After meet~1, SVR $R^2=0.025$ for men and $-0.006$ for women. Latent slope SD should not be confused with individual certainty: empirical-Bayes reliability is low in short seasons, so broad fitted heterogeneity can coexist with weak athlete-level identification.

The $15$--$21$\,s weather result is the consequence of applying the adopted rule, not a separate estimate. Profile minima near $k=0.0014$ and $H_0=94$ are compatible with the adopted $0.0016/100$ pair because one-SE sets are wide. The prospective repeated-venue test is stronger measurement evidence: factors trained on 2023--2024 modestly lower 2025 seen-course MAE for both sexes, with incomplete but reported coverage. Full-sample field factors remain retrospective reconstruction only.

At the team level, ERO adds no independent construct: it is total starts times Hill evenness, and evenness adds little conditional on volume. The mechanical max-count null, attendance selection, continuous rank and top-five outcomes, and within-team TOST all point to the same boundary. More racing characterizes stronger programs between programs; year-to-year changes within a program do not support a schedule effect.

RQ3 finds that recorded men and women race at similar intensity; the observed gender gap reflects fewer recorded women, while non-racing roster members remain unobserved. The sex$\times$race-count interaction on improvement is too imprecise to claim a shared or different slope (Section~\ref{sec:rq3}). Most athletes still race few meets (only about one in ten take $4+$ races), so access to competition shapes who accumulates racing experience.

\noindent\textbf{Transfer to varsity settings:}
NCAA, NAIA, and NJCAA programs often build ladders of regular-season meets plus conference and regional championships, and roster caps mean not every athlete gets the same race count. Our club associations give those settings a large empirical baseline where public exports remain scarce, but eligibility rules and championship incentives differ, so transfer should stay cautious. We do not claim that giving every athlete four races causes better placement.

\noindent\textbf{Practical reading for coaches:}
Use Standardized times to compare meets, while recognizing that late-season meet composition still limits fitness interpretation. Do not triage an athlete from the opener: the race-1-only forecast is near noise. More starts and larger rosters characterize successful programs, but the within-team estimates do not justify adding races as an intervention. Prospective schedule experiments remain necessary.

\section{Limitations}

Finish times omit within-race tactics, mid-race injury, and intentional conservation. We lack training logs, illness records, and menstrual-cycle information. Mixed-effects and ablation analyses (Appendix) mitigate some confounding but cannot replace randomized or quasi-experimental designs. The comprehensive era covers three seasons (2023--2025); 2023 is the first and is thinner on the calendar (no August meets; sparse late-season meets), so primary 2023$\to$2024 temporal splits train on an atypical year (we therefore check 2024$\to$2025). Grade year and pre-2023 training history are unobserved. Improvement analyses condition on $\ge 2$ races, so one-race athletes ($43.1\%$ of regular-season athlete--seasons) are selected out. They are slower, later, on smaller rosters, and about half as likely to return (Appendix~\ref{app:analytic_sample}). We deliberately do \emph{not} impute experience from the historical NRCD era: that export is fragmentary ($43.9\%$ of historical athlete--years show a single recorded race; weather coverage on historical course-detail rows is $\sim18\%$), so race counts would systematically undercount true volume, often by a large fraction. Standardization formulas and package defaults are summarized in Appendix~\ref{app:std_formulas} (weather $k$ adopted from \cite{MaxPerformanceRunning}, not estimated here). Race counts are NRCD-recorded starts; varsity opens and unrecorded meets may undercount true exposure. Our club field factor is a LACCTiC-motivated sensitivity without track-PR initialization; both our EM and standard LACCTiC hold athlete fitness fixed within the fit window unless a time-varying $f_i(t)$ is specified. Roster size, ERO, Mundlak within--between, and within-team first differences show the nationals association is entangled with between-program selection (Figure~\ref{fig:team_dag}).

\section{Ethics}
We use the public de-identified NRCD export (v 2.1.0) \cite{nrcddataset2026,karr2026nrcd}. The dataset's resource paper follows proper IRB approval for research of this scope.

\section{Conclusion}
Public results support comparable times and program-level description, but seasonal improvement is only weakly identified once residual meet difficulty is admitted. Repeating-venue models and prospective course factors provide the constructive path. At nationals, successful programs race more, yet that link is between programs and entangled with roster size rather than a within-team schedule effect. Racing to the starting line is therefore about measured opportunity and what the data can identify rather than an empirical claim that assigning more races causes better times or higher placement.

\bibliographystyle{dcu}
\bibliography{sample-base}

\clearpage

\appendix
\section*{Appendix}

\section{Related work}
\label{app:related_work}

This appendix expands the Related Work section of the main text. The five subsections below retain the full literature review.

Five strands of running and sports analytics research situate this study. Individual performance modeling predicts runners' outcomes across distances, contexts, and time. Work on environmental and course factors adjusts for outside conditions so that heterogeneous events can be compared fairly. Machine learning in sports analytics supplies predictive and interpretive models that can inform coaching decisions. Studies of gender parity document the gap between women's and men's representation in sport research. Finally, work on team-level scheduling and competition frequency asks how participation and race planning affect performance. The subsections below expand on each strand and position our contributions.

\subsection{Individual Performance Modeling}
Performance analytics research has focused on predicting individual outcomes based on past results or known physiological/biomechanical variables. Heart-rate variability (HRV), oxygen consumption, and muscle activation, combined with contextual and psychological factors, allow for better prediction of athletic performance across sports \cite{jianjun2025predictive}.  Rich, multidimensional data can increase model performance. Yet this data is difficult to collect on a large scale, since people need to be studied on an individual basis.

Therefore, predicting race performance from limited measurements is common practice. Low-rank matrix completion approaches and individualized power-law models demonstrate strong cross-distance prediction when large historical datasets are available, showing that individual performance structure can be captured compactly and with high accuracy \cite{blythe2016prediction}. Additionally, work on marathon racing seeks to account for distance, elevation, and runner type (front-pack vs. mid-pack), expanding predictive scope beyond simple power laws \cite{dash2024win}.  Additionally, there has been initial prediction improvement studies of track and field and cross country distances within the Stanford running club \cite{millettpredicting}.  This research stems from the Riegel Race Time Prediction Formula, which converts race times across distances \cite{riegel1981athletic}.

\subsection{Environmental and Course Factors}
The Riegel Race Time Prediction Formula  \cite{riegel1981athletic} enables time adjustment for short or long courses. Yet it does not correct for elevation, weather, or air quality. Air pollution has been shown to correlate with slower race times, while temperature and humidity systematically alter endurance performance \cite{cusick2023impact}. Temperature--dew point adjustments used here follow the NRCD resource paper \citep{karr2026nrcd}; practitioner summaries of heat effects include \cite{MaxPerformanceRunning}. Additionally, trail and ultra-running research has introduced terrain-specific difficulty coefficients and pacing variability measures for better real-world prediction \cite{gutierrez2025real}. Cross-country analyses show that rainfall, mud, and course geometry significantly affect finish times \cite{wilson2024assessing}. A complementary tradition estimates course difficulty from the field itself: LACCTiC \citep{mazaheri2026lacctic} iteratively fits multiplicative meet factors from overlapping athletes' times (initialized with track fitness) to produce 5{,}000\,m-equivalent ratings and race simulations, without taking weather or elevation as inputs. Our analyses use NRCD's metadata Standardized path as primary and treat a transparent club LACCTiC-style factor as a sensitivity check, including whether leftover field hardness still tracks temperature after Standardized (Section~\ref{sec:field_relative}).

\subsection{Machine Learning in Sports Analytics}
There has been a rapid growth in machine learning (ML) applications in sports analytics \cite{pietraszewski2025role}. Endurance-specific reviews note that nonlinear models regularly outperform linear baselines but remain constrained by data quality, limited samples, and external-validity issues \cite{rajvsp2025role}. This is due to certain variables such as injuries not being accounted.

ML models can predict recovery, HRV variation, and fatigue response across long-term training with high accuracy\cite{rothschild2024predicting}. Longitudinal work on elite sprinters shows that ML performance improves substantially with richer contextual features \cite{bartosz2025longitudinal}. Meanwhile, explainable-AI frameworks highlight critical interpretability-accuracy tradeoffs limiting coaching adoption \cite{kranzinger2025scoping}.

\subsection{Lack of Gender Parity}
A persistent limitation in sports science research is the underrepresentation of female athletes. Many prior analyses focus exclusively on men or fail to report gender-specific findings, limiting generalizability. Across 1,826 studies on performance supplements, only 23\% of the participants are women and only 14\% of the studies attempt to define a menstrual cycle.
\cite{james2023underrepresentation}. There is also an underrepresentation of women in exercise science, exercise medicine, and physiology research \cite{anderson2023under, smith2022auditing}.

Imbalance means that normative assumptions about performance improvement, fatigue accumulation, and optimal workload are disproportionately derived from male athletes. Moreover, physiological and biomechanical research documents important sex-specific differences in endurance running, including joint kinematics, fatigue response, substrate utilization, and injury risk \cite{besson2022sex,xie2022sex,hunter2023biological}, highlighting that male-derived models may not generalize accurately to women. In practice, this underrepresentation limits the validity of predictive models and training guidelines for female runners.

\subsection{Team-Level Scheduling and Competition Frequency}

Research on competition frequency shows that dense scheduling affects recovery and performance. In professional soccer, fixture-congestion studies demonstrate that short recovery windows elevate injury risk and reduce high-intensity performance, underscoring how cumulative fatigue shapes season-long outcomes \cite{page2023effects,howle2020injury}. While these findings originate outside endurance running, the underlying physiological mechanisms generalize across sport types.

Endurance-focused research similarly indicates that performance varies with workload and seasonal timing. Trail-running studies report that fitness metrics fluctuate with the accumulation of training and racing demands \cite{matos2021variations}, whereas collegiate distance-running studies show that early-season stress and insufficient recovery can impair later-season performance \cite{lundstrom2025pre}. Despite this, scheduling practices in high school and collegiate cross country remain largely unstudied, and race frequency is often set by tradition rather than evidence.

\section{Analysis population: comprehensive-era club / NIRCA XC}
\label{app:analysis_population}

All numerical claims in this paper are for one fixed population: \textbf{comprehensive-era Cross Country in NRCD}, meet dates on or after 2023-08-01 through the 2025 season in export v2.1.0 (23{,}360 results; 7{,}056 athletes; Table~\ref{tab:sample_coverage}). That is the current club sample (predominantly NIRCA programs), not a rolling update, not the historical NRCD era (2004--July 2023), and not NCAA/NAIA/NJCAA varsity bulk results.

We keep this cut for three reasons. First, only the comprehensive era has near-complete course and weather metadata ($99.9\%$), which is required for Standardized times and for the weather-path identity. Historical XC weather coverage is $27.2\%$; we do not impute it. Second, three seasons already bound the temporal tests (train 2023$\to$test 2024 as primary; 2025 as secondary). Adding sparse historical years would inflate $n$ while breaking the environmental finish-time adjustment. Third, NRCD is a club database. Club athletes overlap with but are not identical to varsity NCAA/NAIA/NJCAA populations; transfer of schedule associations should stay cautious (Section~\ref{sec:discussion}).

Meet-level \texttt{nirca\_only} flags NIRCA-only invitationals versus open club fields; we do \emph{not} drop open meets. The unit of the sample is club teams and athletes competing in this era, including open races those teams enter. Future NRCD seasons can extend the same inclusion rule; until then, results should be read as 2023--2025 club XC, not as a discovered law of collegiate cross country.

\begin{table}[htbp]
\centering
\begin{tabular}{llrrr}
\toprule
Year & Gender & Results & Athletes & Meets \\
\midrule
2023 & Women & 2{,}601 & 1{,}153 & 61 \\
2023 & Men & 4{,}595 & 1{,}927 & 73 \\
2024 & Women & 2{,}874 & 1{,}234 & 91 \\
2024 & Men & 4{,}960 & 2{,}115 & 101 \\
2025 & Women & 3{,}180 & 1{,}382 & 96 \\
2025 & Men & 5{,}150 & 2{,}091 & 104 \\
\bottomrule
\end{tabular}
\caption{Comprehensive-era Cross Country counts by season year and gender.}
\label{tab:sample_year_gender}
\end{table}

\section{Analytic-sample selection ($\ge 1$ vs.\ $\ge 2$ races)}
\label{app:analytic_sample}

\begin{figure}[htbp]
\centering
\includegraphics[width=0.95\linewidth]{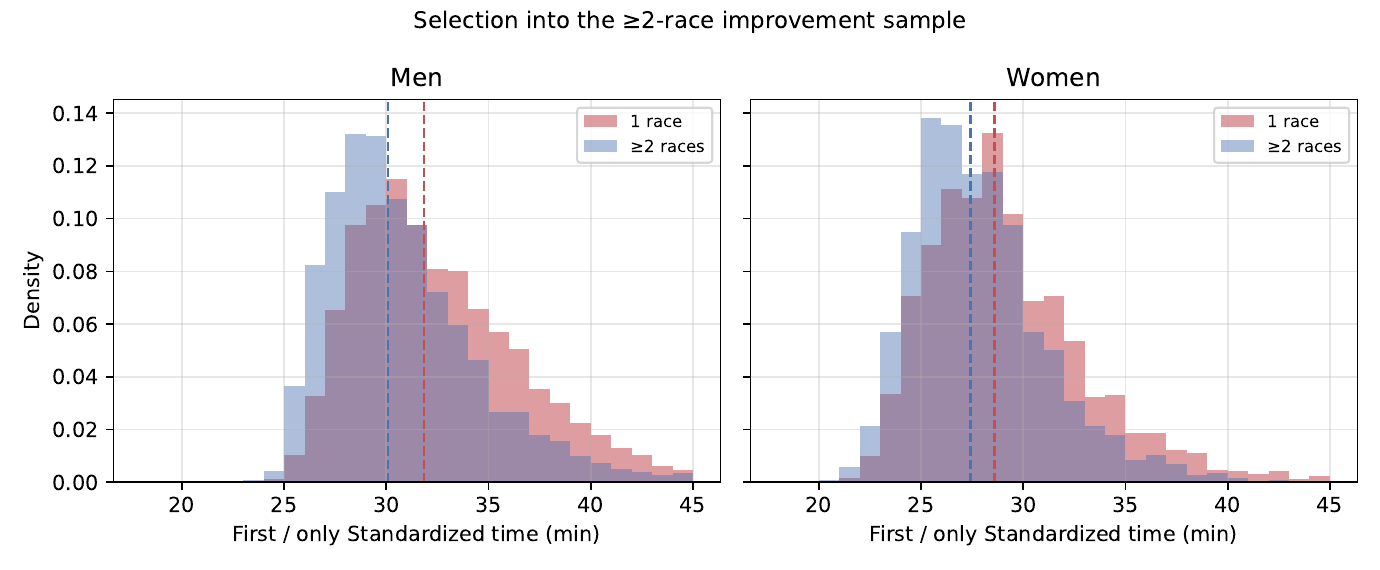}
\caption{Density of first (or only) Standardized time for one-race vs.\ $\ge 2$-race athlete--seasons. Dashed lines are group medians. The improvement sample is shifted toward faster openers.}
\label{fig:selection_first_times}
\end{figure}

Improvement and forecasting analyses require $\ge 2$ regular-season races spanning $\ge 7$ days, so they omit $43.1\%$ of athlete--seasons. Table~\ref{tab:analytic_sample} compares one-race athlete--seasons with those that enter the $\Delta$ sample. One-race athletes are slower, on smaller clubs, start later in the calendar, and are about half as likely to appear the next year. About $32\%$ of one-race seasons occur on or after 15 October. The $\Delta$ sample is therefore selected toward earlier, faster, more persistent athletes on larger rosters: exactly the selection the improvement estimand cannot see (Figure~\ref{fig:selection_first_times}).

\begin{table}[htbp]
\centering
\begin{adjustbox}{max width=\linewidth}
\begin{tabular}{llrrrrr}
\toprule
Gender & Contrast & $n_1$ & $n_{\ge 2}$ & Mean 1-race & Mean $\ge 2$ & $d$ \\
\midrule
Men & First Standardized time (s) & 2545 & 3563 & 1970 & 1862 & $0.39$ \\
Men & Roster size & 2545 & 3563 & 28.0 & 35.1 & $-0.37$ \\
Men & First-race day of year & 2545 & 3563 & 277 & 263 & $1.09$ \\
Men & Next-year return & 1707 & 2321 & $26.8\%$ & $55.3\%$ & -- \\\
Women & First Standardized time (s) & 1704 & 2049 & 1756 & 1675 & $0.35$ \\
Women & Roster size & 1704 & 2049 & 19.3 & 23.9 & $-0.31$ \\
Women & First-race day of year & 1704 & 2049 & 276 & 263 & $1.01$ \\
Women & Next-year return & 1076 & 1303 & $26.5\%$ & $52.6\%$ & -- \\\
\bottomrule
\end{tabular}
\end{adjustbox}
\caption{One-race vs.\ $\ge 2$-race athlete--seasons (nationals excluded, Standardized times; $n_1=4{,}249$, $n_{\ge 2}=5{,}612$). Return rates use 2023--2024 seasons only.}
\label{tab:analytic_sample}
\end{table}

\FloatBarrier

\section{Trajectory identification and early-forecast sensitivity}
\label{app:trajectory_identification}

\noindent\textbf{Crossed meet effects and composition.}
The sparse crossed model adds meet intercepts to athlete--season intercepts and slopes. Its objective is residual sum of squares plus separate ridge penalties; larger meet penalties shrink meet effects more strongly toward zero. Across penalties $0.1,1,10,100$, men's week estimates are $+1.34,-0.50,-3.95,-5.48$\,s/week and women's are $-3.97,-4.31,-4.93,-5.84$. Penalty 10 is a reporting anchor, not a fit-selected optimum. Excluding all championship-like meets, including nationals, and including nationals with an indicator produce the sensitivity values reported in Section~\ref{sec:hier_results}. The championship flag matches meet names containing national(s), regional(s), championship(s), conference, district, or state.

The meet-cluster bootstrap resamples meets, demeans within athlete--season, and refits a meet-ridge model. Its estimates are $-1.68$ $[-5.35,1.97]$\,s/week for men and $-1.31$ $[-4.47,1.94]$ for women (399 draws; seed 20260924). Inverse continuation weights shift them to $-2.08$ and $-1.18$, respectively, but cannot identify unrecorded races. Grade availability is only about $37\%$ overall and is especially sparse late; field size and championship composition also rise strongly with week.

\noindent\textbf{Level-2 and reliability details.}
The prospective descriptive model interacts week with opening Standardized time (per 60\,s), roster size (per 10 athletes), and first-race calendar timing. Opening-time interactions are $-3.79$\,s/week for men and $-2.37$ for women; roster interactions are $-1.51$ and $-0.23$. First-race timing intervals include zero. Random-slope variance rises after adding these terms, so they are not interpreted as explaining heterogeneity. Approximate empirical-Bayes reliability by race count and the retrospective schedule-span/race-count specification are preserved in the generated trajectory artifacts.

\noindent\textbf{Forecast provenance.}
For 2023$\to$2024, the meet-1-only SVR uses opening time and yields $R^2=0.025$ (men) and $-0.006$ (women). Ridge gives $0.005$ and $-0.011$. Meet-2 models add the observed early slope only on seasons eventually known to have at least three races and are secondary. The older six-family full-season suite (Appendix Table~\ref{tab:gender_primary_performance}) uses information accumulated later in the season and is retained for provenance rather than coach-facing interpretation.

\section{Machine-learning specification and feature-exclusion audit}
\label{app:ml_spec}

This appendix reprints the full machine-learning specification notes that were shortened in Section~\ref{sec:ml_estimand}. Nothing below is omitted from that discussion.

\noindent\textbf{Relation to the resource paper's illustrative $R^2$:}
The prior NRCD resource paper \citep{karr2026nrcd} introduced the dataset and, as an illustrative standardization check, reported gender-stratified Random Forest / Gradient Boosting test $R^2$ of roughly $0.5$--$0.74$ under Converted Only vs.\ Standardized labels. That analysis is first in the publication sequence and remains valid for its stated purpose: a \emph{paired} demonstration that Converted Only leaves environmental structure in the label ($\Delta R^2$ favoring Converted), not a claim of out-of-year forecasting skill. Absolute $R^2$ in that demo is high by construction. Its features include within-season summaries built from the result times under test, including $\mathrm{last\_time}$, while the label is $(t_{L}-t_{1})/\Delta$, so a flexible model can nearly reconstruct the outcome; even without $\mathrm{last\_time}$, full-season $\mathrm{best\_time}$ equals the final race for athletes whose peak is last. The present paper builds on that resource by asking the forecasting question the demo did not: with the same improvement-rate estimand, but the outcome race excluded from all predictors and rows defined as athlete--seasons, held-out $R^2$ falls to $\approx 0$ (Table~\ref{tab:gender_primary_performance}). The two results are complementary, not contradictory.

\begin{figure}[htbp]
    \centering
    \includegraphics[width=1\linewidth]{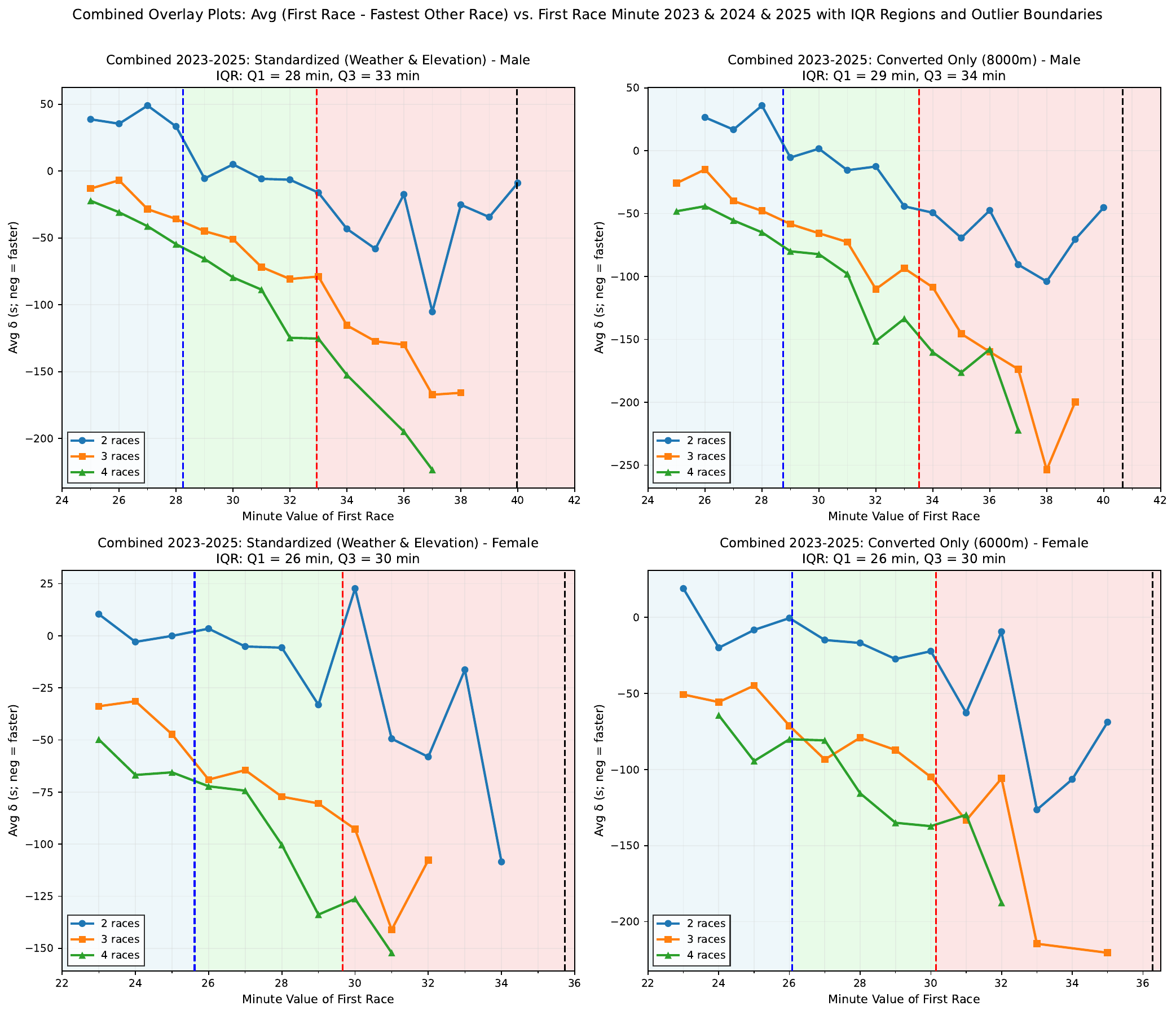}
    \caption{First-to-fastest change $t_{\mathrm{best}}-t_1$ versus first-race time (minutes), combined 2023--2025 (negative $=$ faster). Panels are men/women $\times$ Standardized / Converted Only. Each series is a race-count group (markers in the legend). Points are minute-bin means averaged across years (bin $n\ge 5$, $\ge 2$ years). Within each series, first-race times outside the $1.5\times$IQR fence are dropped before binning. Vertical dashed lines and shaded bands use the \emph{sex-specific} first-race distribution: Q1 (blue; faster starters), the IQR (green), Q3 (red), and the $1.5\times$IQR fence (black). Converted Only is distance-converted without weather/elevation. Compare median first-to-last strata in Table~\ref{tab:dose_response} and Appendix~\ref{app:subgroup}.}
    \label{fig:combined_improvement}
\end{figure}

\noindent\textbf{Primary vs.\ supporting improvement metrics:}
The primary within-season estimand is the hierarchical trajectory in Equation~\ref{eq:hier_trajectory} (every Standardized race; athlete--season random intercept and random slope on week of season). The two-point rate in Equation~\ref{eq:improvement_rate} is a \emph{supporting} coach-facing summary and the target of leakage-controlled SVR forecasts; the descriptive dose--response table (Table~\ref{tab:dose_response}) reports the same signed first-to-last change $\delta=t_L-t_1$ in total seconds (negative $=$ faster; the rate divides by $\Delta_{i,s}$ days). Two further definitions serve only as robustness checks of the two-point summary, not as separate findings:
\begin{itemize}
\item the per-race linear-regression trajectory slope over races $1,\ldots,L-1$ (Appendix~\ref{app:subgroup} subgroup medians and the \texttt{slope} feature), which is less endpoint-sensitive but ignores the final race;
\item first-to-\emph{fastest} change $t_{\mathrm{best}}-t_1$ (Figure~\ref{fig:combined_improvement}; negative $=$ faster), which is bounded above by zero and therefore mechanically larger in magnitude than signed first-to-last $\delta$ when the best race is not last.
\end{itemize}
The hierarchical slope and the two-point summaries agree that the field gets faster; they disagree in magnitude because the two-point rate puts full weight on noisy endpoints (and inherits Galton inflation when slow openers rebound), whereas the hierarchical model uses every race and shrinks extremes. Where two-point robustness checks diverge among themselves, as with the dose--response peak at three to four races versus monotone-looking subgroup slopes, the difference comes from endpoint choice rather than a substantive contradiction with the hierarchical result.

\noindent\textbf{Hyperparameter policy:}
Primary reported learner is RBF SVR ($C=1$, $\gamma=\texttt{scale}$) with \texttt{StandardScaler} in the pipeline. Other families use stock scikit-learn defaults with only \texttt{random\_state=42} (Random Forest / Gradient Boosting: $n_{\mathrm{estimators}}=100$; Ridge $\alpha=1$; Lasso $\alpha=0.01$). There is no nested CV or grid search and no test-set model selection. Default trees going well below $R^2=0$ is expected overfit of a weak 2023 label and is not a tuned null claim about forests. Seed 42 for bootstrap and sklearn RNGs.

\noindent\textbf{Feature exclusion policy (full classes):}
We maintain a single feature-exclusion policy with one exclusion class per feature (no double-counting), then prune near-duplicate \emph{retained} covariates ($|\mathrm{corr}|\ge0.95$) to a compact primary set of 17 within-sex predictors (one representative per collinear cluster: e.g., $\texttt{cv\_time}$ for variability transforms; $\texttt{first\_time}$/$\texttt{best\_time}$ for absolute times). Classes:
\begin{itemize}
\item \textbf{A (endpoint leakage):} $\texttt{last\_time}$, $\texttt{total\_improvement}$, and their aliases ($\texttt{late\_season\_performance}$, $\texttt{progression\_improvement}$).
\item \textbf{B (derived leakage):} $\texttt{improvement\_per\_race}$, $\texttt{improvement\_to\_variability\_ratio}$.
\item \textbf{C (exact duplicate of a retained column):} $\texttt{races\_duration\_ratio}$ ($\equiv\texttt{race\_frequency}$), $\texttt{early\_season\_performance}$ ($\equiv\texttt{first\_time}$).
\item \textbf{D (retained, non-external):} pre-final $\texttt{slope}$, compact absolute times, schedule descriptors (races $1,\ldots,L-1$ or season totals). Not early-season forecasts; for two-race athletes $\texttt{slope}=0$.
\item \textbf{E (within-sex drop):} gender encodings after sex-stratified fitting.
\end{itemize}
Temporal audits (Table~\ref{tab:feature_exclusion}) confirm the policy: add-back of $\texttt{last\_time}$ raises men's $R^2$ by $+0.44$; $\texttt{total\_improvement}$ by $+0.65$; exact duplicates change $R^2$ by $<0.01$. Compact vs.\ legacy-full $\Delta R^2=-0.007$ (men), so redundancy pruning does not manufacture the null. Dropping $\texttt{slope}$ or schedule blocks leaves $R^2\approx0$; early-window specs stay near zero.

\begin{table}[htbp]
\centering
\begin{adjustbox}{max width=\linewidth}
\begin{tabular}{llrrl}
\toprule
Spec & Gender & Test $R^2$ & $\Delta R^2$ vs compact & Verdict \\
\midrule
Compact primary (17 feats) & Men & $0.044$ & -- & baseline \\
Legacy full (26 feats) & Men & $0.051$ & $+0.007$ & prune OK \\
Compact $+$ \texttt{last\_time} & Men & $0.478$ & $+0.434$ & leakage \\
Compact $+$ \texttt{total\_improvement} & Men & $0.696$ & $+0.652$ & leakage \\
Compact $+$ \texttt{races\_duration\_ratio} & Men & $0.041$ & $-0.003$ & duplicate OK \\
Drop \texttt{slope} & Men & $0.035$ & $-0.009$ & null robust \\
Early-window proxy & Men & $0.032$ & $-0.012$ & still $\approx0$ \\
\midrule
Compact primary & Women & $-0.018$ & -- & baseline \\
Legacy full & Women & $0.001$ & $+0.019$ & prune OK \\
Compact $+$ \texttt{last\_time} & Women & $0.368$ & $+0.386$ & leakage \\
Drop \texttt{slope} & Women & $-0.026$ & $-0.008$ & null robust \\
\bottomrule
\end{tabular}
\end{adjustbox}
\caption{Feature-exclusion audit after adopting the compact primary set (SVR; train 2023$\to$test 2024). Leakage add-backs spike $R^2$; redundancy prune vs.\ legacy-full does not.}
\label{tab:feature_exclusion}
\end{table}

\section{Supplementary results figures and tables}
\label{app:supp_results}

Floats below were moved from the main text for length. Cross-references in the main text still point here.

\begin{table}[htbp]
\centering
\begin{adjustbox}{max width=\linewidth}
\begin{tabular}{lllrrr}
\toprule
Split & Gender & Method & Test $n$ & MAE (s) & $R^2$ \\
\midrule
2023$\to$2024 & Men & Persistence & 1167 & $84.2$ & $0.682$ \\
2023$\to$2024 & Men & EB early slope & 1167 & $83.5$ & $0.697$ \\
2023$\to$2024 & Women & Persistence & 670 & $98.1$ & $0.532$ \\
2023$\to$2024 & Women & EB early slope & 670 & $95.7$ & $0.549$ \\
\midrule
2024$\to$2025 & Men & Persistence & 1226 & $108.3$ & $-0.316$ \\
2024$\to$2025 & Men & EB early slope & 1226 & $104.1$ & $-0.206$ \\
2024$\to$2025 & Women & Persistence & 735 & $71.9$ & $0.6400$ \\
2024$\to$2025 & Women & EB early slope & 735 & $77.2$ & $0.6398$ \\
\bottomrule
\end{tabular}
\end{adjustbox}
\caption{Out-of-year last-race forecast from the hierarchical model (predict last Standardized time from the opener). EB uses races before the last when available, else the training-year population slope. Women's 2024$\to$2025 $R^2$ values are nearly identical ($0.6400$ vs.\ $0.6398$) while MAE differs because MAE and $R^2$ (squared error) can rank methods differently. Men's 2024$\to$2025 persistence $R^2=-0.316$ is not a bug: 2025 men improved steeply (Table~\ref{tab:hier_per_year}), so ``last $=$ first'' misses a large mean shift and underperforms predicting the test-year mean. Persistence remains competitive on MAE; slope transfer does not clearly beat it.}
\label{tab:hier_oos}
\end{table}

\begin{table}[htbp]
\centering
\begin{adjustbox}{max width=\linewidth}
\begin{tabular}{llrrrr}
\toprule
Gender & Model & Train $n$ & Test $n$ & Test $R^2$ (95\% CI) & MAE \\
\midrule
\multicolumn{6}{l}{\textit{Primary: train 2023 $\to$ test 2024}} \\
Men & SVR & 1147 & 1164 & $0.044$ [$-0.021$, $0.102$] & 3.66 \\
Men & Lasso & 1147 & 1164 & $0.007$ [$-0.052$, $0.059$] & 3.77 \\
Men & Ridge & 1147 & 1164 & $-0.006$ [$-0.067$, $0.052$] & 3.81 \\
Men & OLS & 1147 & 1164 & $-0.021$ [$-0.086$, $0.036$] & 3.87 \\
Men & Gradient Boosting & 1147 & 1164 & $-0.139$ [$-0.266$, $-0.028$] & 4.05 \\
Men & Random Forest & 1147 & 1164 & $-0.403$ [$-0.619$, $-0.219$] & 4.32 \\
Women & SVR & 631 & 670 & $-0.018$ [$-0.091$, $0.041$] & 4.14 \\
Women & Lasso & 631 & 670 & $-0.125$ [$-0.232$, $-0.024$] & 4.24 \\
Women & Ridge & 631 & 670 & $-0.121$ [$-0.229$, $-0.019$] & 4.24 \\
Women & OLS & 631 & 670 & $-0.125$ [$-0.236$, $-0.021$] & 4.29 \\
Women & Gradient Boosting & 631 & 670 & $-0.136$ [$-0.296$, $-0.006$] & 4.36 \\
Women & Random Forest & 631 & 670 & $-0.284$ [$-0.513$, $-0.093$] & 4.48 \\
\midrule
\multicolumn{6}{l}{\textit{Sensitivity: train 2024 $\to$ test 2025 (SVR only)}} \\
Men & SVR & 1164 & 1220 & $0.056$ [$0.031$, $0.083$] & 3.68 \\
Women & SVR & 670 & 734 & $-0.060$ [$-0.206$, $0.056$] & 3.78 \\
\bottomrule
\end{tabular}
\end{adjustbox}
\caption{Gender-separated Standardized-time model performance on the compact feature set. Primary block: train 2023$\to$test 2024 (2{,}000-bootstrap CIs); SVR is the pre-specified learner for this supporting task, and other families are untuned sensitivities. Neither SVR interval excludes zero. Sensitivity rows train 2024$\to$test 2025 for SVR only. Men's primary-split MAE skill vs.\ the training-mean baseline is $7.6\%$ (women $8.2\%$); after mean-shifting predictions to the test-year mean, $R^2\approx 0.08$ for both sexes. $\mathrm{SER}=0.19$ for men and is non-positive for women.}
\label{tab:gender_primary_performance}
\end{table}

\begin{table}[htbp]
\centering
\begin{adjustbox}{max width=\linewidth}
\begin{tabular}{llrrrr}
\toprule
 & & \multicolumn{4}{c}{Median $\delta=t_L-t_1$ (s; neg.\ $=$ faster)} \\
\cmidrule(lr){3-6}
Gender & Start quartile & 2 & 3 & 4 & $5+$ \\
\midrule
\multicolumn{6}{l}{\textit{A.\ Overall (all starters; $n$ and 95\% bootstrap CI below)}} \\
Men & -- & $-10.9$ & $-17.5$ & $-19.8$ & $-4.7$ \\
 & & {\footnotesize $n{=}1642$} & {\footnotesize $1171$} & {\footnotesize $593$} & {\footnotesize $136$} \\
 & & {\footnotesize $[-16.8,-5.2]$} & {\footnotesize $[-22.5,-12.1]$} & {\footnotesize $[-28.9,-12.2]$} & {\footnotesize $[-24.2,12.5]$} \\
Women & -- & $-13.7$ & $-20.3$ & $-28.8$ & $-20.2$ \\
 & & {\footnotesize $n{=}1014$} & {\footnotesize $633$} & {\footnotesize $317$} & {\footnotesize $72$} \\
 & & {\footnotesize $[-19.3,-8.3]$} & {\footnotesize $[-28.4,-14.6]$} & {\footnotesize $[-39.6,-19.5]$} & {\footnotesize $[-27.5,8.3]$} \\
\midrule
\multicolumn{6}{l}{\textit{B.\ Men by first-race Standardized quartile (Q1 $=$ fastest; $n$ under each median)}} \\
Men & Q1 (fastest) & $20.1$ & $16.4$ & $3.5$ & $23.0$ \\
 & & {\footnotesize $n{=}332$} & {\footnotesize $302$} & {\footnotesize $190$} & {\footnotesize $63$} \\
Men & Q2 & $-7.5$ & $4.5$ & $-15.5$ & $-25.6$ \\
 & & {\footnotesize $n{=}395$} & {\footnotesize $287$} & {\footnotesize $169$} & {\footnotesize $34$} \\
Men & Q3 & $-23.0$ & $-40.9$ & $-56.2$ & $-63.6$ \\
 & & {\footnotesize $n{=}448$} & {\footnotesize $277$} & {\footnotesize $136$} & {\footnotesize $23$} \\
Men & Q4 (slowest) & $-53.7$ & $-89.5$ & $-101.3$ & $-73.6$ \\
 & & {\footnotesize $n{=}467$} & {\footnotesize $305$} & {\footnotesize $98$} & {\footnotesize $16$} \\
\midrule
\multicolumn{6}{l}{\textit{C.\ Women by first-race Standardized quartile}} \\
Women & Q1 (fastest) & $9.6$ & $-2.6$ & $-8.8$ & $12.4$ \\
 & & {\footnotesize $n{=}203$} & {\footnotesize $173$} & {\footnotesize $101$} & {\footnotesize $33$} \\
Women & Q2 & $-10.6$ & $-10.3$ & $-20.1$ & $-26.5$ \\
 & & {\footnotesize $n{=}253$} & {\footnotesize $151$} & {\footnotesize $86$} & {\footnotesize $19$} \\
Women & Q3 & $-17.4$ & $-18.7$ & $-53.2$ & $-23.7$ \\
 & & {\footnotesize $n{=}272$} & {\footnotesize $149$} & {\footnotesize $72$} & {\footnotesize $14$} \\
Women & Q4 (slowest) & $-65.3$ & $-72.9$ & $-108.6$ & -- \\
 & & {\footnotesize $n{=}286$} & {\footnotesize $160$} & {\footnotesize $58$} & {\footnotesize $n{<}10$} \\
\bottomrule
\end{tabular}
\end{adjustbox}
\caption{Median Standardized first-to-last change $\delta=t_L-t_1$ (seconds; negative $=$ faster; nationals excluded). Panel~A: overall by gender and race count with $n$ and 95\% bootstrap CIs. Panels~B--C: gender--year first-race Standardized quartiles on the $\ge 2$-race sample (equal overall $n$ per quartile; unequal within race-count columns because slower starters race fewer times). Women's Q4 $5+$ omitted ($n<10$). First-to-last cell medians are affected by regression to the mean when binning on $t_1$; see Figure~\ref{fig:rtm_vs_ability} for the valid race-2$\to$last check. Cohen's $d$ vs.\ two races (means): men $0.18$/$0.08$/$-0.01$ at $3/4/5+$; women $0.11$/$0.12$/$-0.06$ (unchanged under sign flip).}
\label{tab:dose_response}
\end{table}

\begin{table}[htbp]
\centering
\begin{adjustbox}{max width=\linewidth}
\begin{tabular}{llrrrr}
\toprule
Gender & Method & $n$ & Mean $\delta$ (s) & 95\% CI & \% faster ($\delta<0$) \\
\midrule
Men & Standardized & 3542 & $-26.1$ & $[-32.1,\,-20.5]$ & 57.2 \\
Men & Converted Only & 3542 & $-47.3$ & $[-53.1,\,-41.6]$ & 65.3 \\
Men & Gap (Std.$-$Conv.) & 3542 & $21.2$ & $[20.4,\,22.0]$ & -- \\
\midrule
Women & Standardized & 2036 & $-23.2$ & $[-28.3,\,-18.1]$ & 59.5 \\
Women & Converted Only & 2036 & $-38.6$ & $[-43.8,\,-33.4]$ & 66.8 \\
Women & Gap (Std.$-$Conv.) & 2036 & $15.4$ & $[14.5,\,16.3]$ & -- \\
\bottomrule
\end{tabular}
\end{adjustbox}
\caption{Weather inflation: paired first-to-last change $\delta=t_L-t_1$ under Standardized vs.\ Converted Only (negative $=$ faster; 95\% bootstrap CIs). Gap $=$ Std.\ minus Conv.\ is positive because Converted is more negative (overstates improvement by $15$--$21$\,s). Gaps reject equality by one-sided Wilcoxon at $p\ll 0.001$.}
\label{tab:weather_inflation}
\end{table}

\begin{figure}[htbp]
\centering
\includegraphics[width=0.95\linewidth]{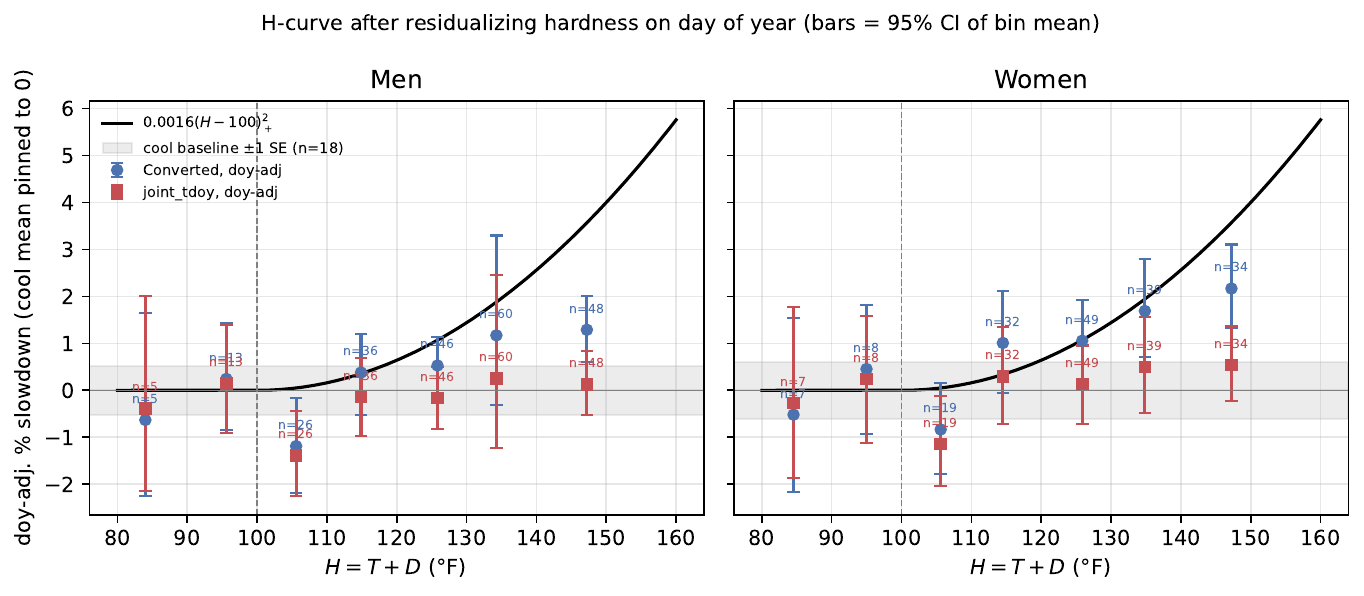}
\caption{Day-of-year--adjusted ability-netted Converted meet hardness (and joint\_tdoy leftover hardness) by $H=T+D$ bin, overlaid on $0.0016(H-100)_+^{2}$. Hardness is $1/\alpha$ from the field EM on Converted times; $\log$ hardness is residualized on day of year and the cool-meet ($H\le 100$) mean residual is pinned to $0$. Bars are bootstrap 95\% CIs of the bin mean (not within-bin meet spread); $n$ per bin is annotated. The gray band is $\pm 1$\,SE of the cool baseline ($n=18$ men / $15$ women). Without the doy control the Converted curve is calendar-confounded; after it, Converted sits nearer the heuristic and joint\_tdoy is flatter, matching joint\_tdoy's smaller linear heat coefficient.}
\label{fig:heuristic_h_curve}
\end{figure}

\begin{table}[htbp]
\centering
\begin{adjustbox}{max width=\linewidth}
\begin{tabular}{lrrrrr}
\toprule
Gender & Interval & $n$ & Mean $\Delta$ (s) & 95\% CI & \% faster \\
\midrule
Men & 2023$\to$2024 & 385 & $+5.3$ & $[-7.4,\,18.2]$ & 56.9 \\
Men & 2024$\to$2025 & 425 & $-4.4$ & $[-17.9,\,10.0]$ & 54.4 \\
Men & 2023$\to$2025 & 243 & $-3.9$ & $[-20.2,\,13.3]$ & 54.7 \\
Women & 2023$\to$2024 & 210 & $-1.0$ & $[-14.7,\,13.0]$ & 52.9 \\
Women & 2024$\to$2025 & 211 & $-8.8$ & $[-25.7,\,8.8]$ & 55.9 \\
Women & 2023$\to$2025 & 110 & $-21.1$ & $[-45.8,\,6.3]$ & 62.7 \\
\bottomrule
\end{tabular}
\end{adjustbox}
\caption{Mean change in best Standardized time (seconds; negative $=$ faster; 95\% bootstrap CIs) among race-count--consistent athletes. Rows are pairwise adjacent-season panels with $|\Delta$ races$|\le 1$ (interval-specific $n$), not a single multi-interval panel with one $n$ per gender; signs can flip across intervals under that filter. Every CI includes zero. Primary claim remains the null multi-season forecast on the 2023$\to$2024 split.}
\label{tab:rq2_best_time}
\end{table}

\begin{table}[htbp]
\centering
\begin{adjustbox}{max width=\linewidth}
\begin{tabular}{llrrr}
\toprule
Gender & Prior races & $n$ & Retention & 95\% CI \\
\midrule
Men & 2 & 1083 & 0.449 & $[0.419,\,0.479]$ \\
Men & 3 & 765 & 0.603 & $[0.567,\,0.639]$ \\
Men & 4 & 395 & 0.716 & $[0.668,\,0.759]$ \\
Men & 5+ & 70 & 0.671 & $[0.557,\,0.786]$ \\
Men & All ($\ge 2$) & 2313 & 0.552 & $[0.533,\,0.573]$ \\
\midrule
Women & 2 & 630 & 0.414 & $[0.375,\,0.452]$ \\
Women & 3 & 420 & 0.621 & $[0.576,\,0.667]$ \\
Women & 4 & 209 & 0.612 & $[0.541,\,0.679]$ \\
Women & 5+ & 38 & 0.789 & $[0.658,\,0.921]$ \\
Women & All ($\ge 2$) & 1297 & 0.524 & $[0.497,\,0.552]$ \\
\bottomrule
\end{tabular}
\end{adjustbox}
\caption{Next-season retention by prior regular-season race count (2023--2024 athlete--seasons with $\ge 2$ races).}
\label{tab:retention}
\end{table}

\begin{table}[ht]
    \centering
    \begin{adjustbox}{max width=\linewidth}
    \begin{tabular}{lrrrrrr}
        \toprule
        Category & 2023 M & 2024 M & 2025 M & 2023 W & 2024 W & 2025 W \\
        \midrule
        Unique athletes & 1920 & 2119 & 2086 & 1149 & 1234 & 1373 \\
        Total races & 3906 & 4229 & 4316 & 2206 & 2433 & 2652 \\
        Avg races / athlete & 2.03 & 2.00 & 2.07 & 1.92 & 1.97 & 1.93 \\
        \bottomrule
    \end{tabular}
    \end{adjustbox}
    \caption{Comprehensive-era Cross Country participation by year and gender (regular season includes all non-filtered meets used in RQ3). Counts are descriptive; equal gender shares are not a substantive null. Race-count bins are in Table~\ref{tab:2023_gender_race_distribution}.}
    \label{tab:total_unique_results}
\end{table}

\begin{figure}[htbp]
\centering
\includegraphics[width=0.95\linewidth]{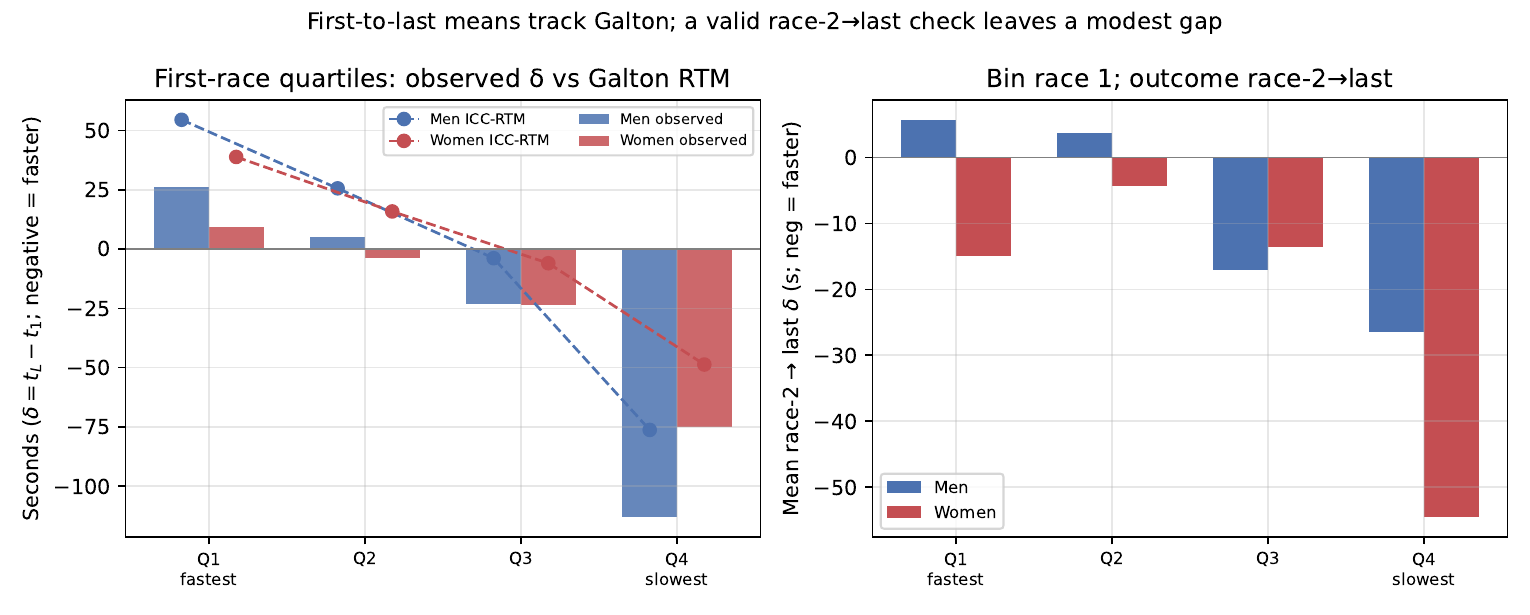}
\caption{Left: mean first-to-last $\delta=t_L-t_1$ (negative $=$ faster) by first-race quartile tracks the Galton/ICC prediction $(1-\mathrm{ICC})(\mu-t_1)$. Right: valid check bins on race~1 and measures race-2$\to$last $\delta$ (bin independent of both endpoints); the Q4--Q1 gap is modest (men $32$\,s, $d=0.31$; women $40$\,s, $d=0.29$).}
\label{fig:rtm_vs_ability}
\end{figure}


\begin{figure}[htbp]
\centering
\includegraphics[width=0.92\linewidth]{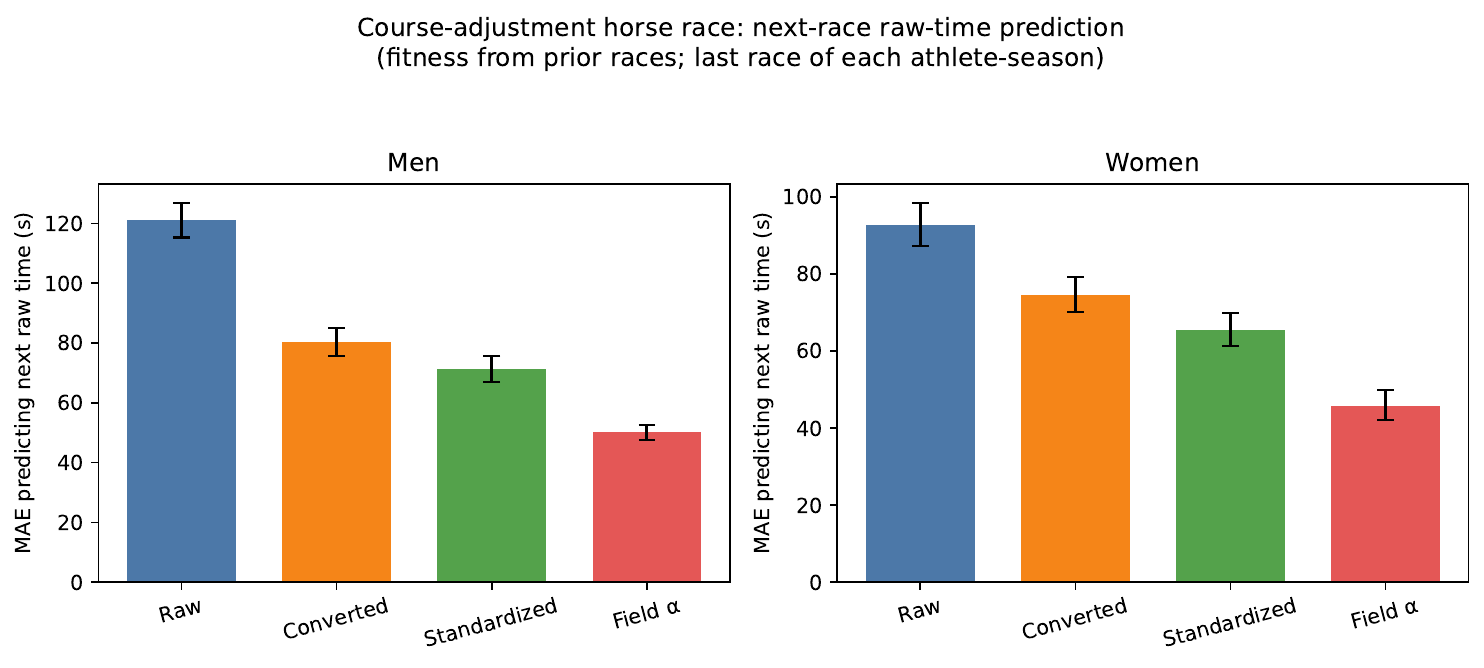}
\caption{Next-race raw-time prediction MAE (last race of each athlete--season; 95\% bootstrap CIs) under Raw, Converted Only, Standardized, and field-$\alpha$ ability. Field $\alpha$ beats Standardized on next-raw MAE and race order; Standardized remains preferred for weather-adjusted improvement; Std-then-$\alpha$ is the best race-prediction hybrid (Table~\ref{tab:course_method_horse}; Figure~\ref{fig:std_lacctic_race}).}
\label{fig:next_race_mae}
\end{figure}

\begin{table}[ht!]
    \centering
    \begin{adjustbox}{max width=\linewidth}
    \begin{tabular}{lcccccc}
        \toprule
        Group & Teams & 4+ teams & Top-15 with 4+ & RR & $p$ ($\chi^2$) & Bonf.\ sig. \\
        \midrule
        2023 Men & 92 & 35 (38\%) & 7/14 (50\%) & 1.63 & 0.483 & No \\
        2023 Women & 83 & 25 (30\%) & 7/15 (47\%) & 2.03 & 0.218 & No \\
        2024 Men & 104 & 36 (35\%) & 8/15 (53\%) & 2.16 & 0.176 & No \\
        2024 Women & 91 & 28 (31\%) & 9/15 (60\%) & 3.38 & 0.017 & No \\
        2025 Men & 106 & 39 (37\%) & 6/15 (40\%) & 1.15 & 1.000 & No \\
        2025 Women & 97 & 33 (34\%) & 9/15 (60\%) & 2.91 & 0.044 & No \\
        \bottomrule
    \end{tabular}
    \end{adjustbox}
    \caption{Per-cell association: $\ge 1$ athlete with $4+$ races vs.\ top-15 finish. Bonferroni $\alpha=0.0083$; no cell survives. Robust pooled/continuous/depth results are in Table~\ref{tab:team_robust}.}
    \label{tab:per_nats}
\end{table}

\begin{table}[htbp]
\centering
\begin{adjustbox}{max width=\linewidth}
\begin{tabular}{llrr}
\toprule
Split & Gender & $n_{\text{test}}$ & Primary SVR $R^2$ (95\% CI) \\
\midrule
2023 $\to$ 2024 & Men & 198 & $-0.09$ $[-0.40,0.05]$ \\
2023 $\to$ 2024 & Women & 90 & $-0.08$ $[-0.42,0.03]$ \\
2023 $\to$ 2025 & Men & 197 & $-0.07$ $[-0.35,0.08]$ \\
2023 $\to$ 2025 & Women & 77 & $-0.09$ $[-0.46,0.07]$ \\
2023+2024 $\to$ 2025 & Men & 197 & $-0.03$ $[-0.28,0.12]$ \\
2023+2024 $\to$ 2025 & Women & 77 & $-0.03$ $[-0.40,0.15]$ \\
\bottomrule
\end{tabular}
\end{adjustbox}
\caption{Multi-season held-out performance by temporal split and sex for the pre-specified SVR with compact leakage-controlled features. The primary 2023$\rightarrow$2024 split is non-positive. Unselected sensitivity families (e.g., men's 2025 Lasso) can look large but are not used for model selection; athlete overlap across years is not treated as leakage.}
\label{tab:rq2_splits}
\end{table}

\begin{table}[ht]
    \centering
    \begin{adjustbox}{max width=\linewidth}
    \begin{tabular}{lrrrrrr}
        \toprule
        Race count & 2023 M & 2024 M & 2025 M & 2023 W & 2024 W & 2025 W \\
        \midrule
        1 & 770 (40.1\%) & 951 (44.9\%) & 847 (40.6\%) & 517 (45.0\%) & 567 (45.9\%) & 628 (45.7\%) \\
        2 & 543 (28.3\%) & 545 (25.7\%) & 577 (27.7\%) & 319 (27.8\%) & 313 (25.4\%) & 391 (28.5\%) \\
        3 & 396 (20.6\%) & 369 (17.4\%) & 402 (19.3\%) & 206 (17.9\%) & 214 (17.3\%) & 212 (15.4\%) \\
        4+ & 211 (11.0\%) & 254 (12.0\%) & 260 (12.5\%) & 107 (9.3\%) & 140 (11.3\%) & 142 (10.3\%) \\
        \bottomrule
    \end{tabular}
    \end{adjustbox}
    \caption{Comprehensive-era race-count distribution by year and gender (athlete--seasons; nationals excluded). Bin totals can differ slightly from Table~\ref{tab:analytic_sample} when a season lacks a usable Standardized time.}
    \label{tab:2023_gender_race_distribution}
\end{table}

\FloatBarrier

\section{IQR-Based Subgroup Analysis by Fitness Level}
\label{app:subgroup}

To explore how initial fitness level impacts improvement trajectories, we categorized runners into ten percentile groups based on their first race time. Within each group we report the median of athlete-level ordinary-least-squares slopes of Standardized time on race index (races $1,\ldots,L$; seconds per race-step; negative $=$ getting faster). Positive medians in the fastest groups therefore mean those athletes slowed on average. These subgroup slopes are exploratory and uncorrected for multiplicity.

Across sex\,$\times$\,year\,$\times$\,standardization panels, slower initial times associate with larger first-to-last gains, but that pattern is affected by regression to the mean when $\Delta$ shares the binning race (Section~\ref{sec:dose_response}; Figure~\ref{fig:rtm_vs_ability}). We show one representative panel (men, 2023, Standardized; Table~\ref{tab:2023_stand}). Converted Only shows uniformly larger apparent gains than Standardized. The continuous association is in Figure~\ref{fig:combined_improvement}.

\subsubsection*{10 Subgroups within 1.5 × IQR of First Race Times}

\begin{table}[htbp]
\renewcommand{\arraystretch}{1.1}
\centering
\begin{tabular}{c c c c c}
\toprule
Subgroup& Time Range & Median Slope& CV& Num Athletes \\
\midrule
0--10\%   & 23:44.7--27:02.0 &  1.27 & 2.32  & 111 \\
    10--20\%  & 27:02.1--27:56.4 &  0.80 & 3.10  & 111 \\
    20--30\%  & 27:56.6--28:39.0 &  0.94 & 3.64  & 111 \\
    30--40\%  & 28:39.0--29:21.6 &  0.26 & 6.57  & 111 \\
    40--50\%  & 29:22.5--30:07.1 &  0.26 & 9.80  & 111 \\
    50--60\%  & 30:07.2--30:55.3 &  0.28 & 7.17  & 111 \\
    60--70\%  & 30:55.9--31:53.9 & -0.67 & --$^{\dagger}$ & 111 \\
    70--80\%  & 31:54.2--33:08.9 & -1.18 & 6.57  & 111 \\
    80--90\%  & 33:09.0--34:50.0 & -1.87 & 3.62  & 111 \\
    90--100\% & 34:50.4--39:12.5 & -2.65 & 3.06  & 111 \\
\bottomrule
\end{tabular}
\caption{Representative subgroup panel: men, 2023, Standardized. Median within-athlete OLS slope of Standardized time on race index (s/race-step; negative $=$ getting faster). Positive medians in fast groups mean slowing. $^{\dagger}$CV omitted when the median is near zero (unstable ratio).}
\label{tab:2023_stand}
\end{table}

\FloatBarrier
\section{Model Diagnostic Assessments}
\label{app:diagnostics}

Residual diagnostics for pooled models and a pooled-gender SVR ($R^2=0.041$) are omitted for brevity; neither rescues the weak temporal $R^2$. Gender-separated models remain primary.

\section{Individual-Level Model Robustness, Sensitivity, and Mixed Effects (RQ1)}
\label{app:rq1_ml_supplement}

This section supplements the individual-level models (train 2023, test 2024 unless noted). Because baseline $R^2$ is non-positive, ablations diagnose instability rather than identify useful predictors.

\subsection{Feature ablation and sensitivity}
\label{app:rq1_ablation}

A fixed random forest ($n_{\mathrm{estimators}}=100$) on gender-specific subsets (endpoint time excluded) yields baseline test $R^2=-0.419$ (men) and $-0.306$ (women). Single-feature removals change $R^2$ by at most $\approx0.02$; dropping the top-10 importance features together still leaves $R^2<0$ (men $-0.247$, women $-0.292$). Sensitivity sweeps over clipping ($\pm20$/$\pm100$), squared terms, absolute time markers, and forest size keep men's primary-split $R^2$ in roughly $[-0.48,-0.37]$ and women's non-positive.

\subsection{Learning curves and outcome-trimming sensitivity}
\label{app:prediction_diagnostics}

Diagnostics below use the same compact primary feature set as the main text (17 within-sex predictors).

To distinguish a low-signal result from one caused only by limited training size, we fit SVR, Random Forest, and Gradient Boosting on fixed fractions of the 2023 gender-specific training set and always evaluated on the full 2024 test set. At each 20--80\% fraction we drew 20 subsets with seeds 42--61; the full-data fit is unique. Table~\ref{tab:learning_curves} shows that SVR moves only from mean $R^2=0.023$ to $0.044$ for men and stays negative for women between 20\% and 100\% of training data, with curves effectively flat from 80\% to 100\%. Gradient Boosting and Random Forest remain negative at full size. Learning curves cannot prove an irreducible ceiling, but their late flattening makes ``the same features merely need modestly more observations'' an insufficient explanation for the null result.

\begin{table}[htbp]
\centering
\begin{adjustbox}{max width=\linewidth}
\begin{tabular}{llrrrrr}
\toprule
 & & \multicolumn{5}{c}{Mean 2024 test $R^2$ by \% of 2023 training data} \\
\cmidrule(lr){3-7}
Gender & Model & 20\% & 40\% & 60\% & 80\% & 100\% \\
\midrule
Men & SVR & $0.023$ & $0.033$ & $0.050$ & $0.046$ & $0.044$ \\
Men & Gradient Boosting & $-0.477$ & $-0.278$ & $-0.198$ & $-0.184$ & $-0.139$ \\
Men & Random Forest & $-0.392$ & $-0.368$ & $-0.348$ & $-0.374$ & $-0.403$ \\
\midrule
Women & SVR & $-0.022$ & $-0.019$ & $-0.010$ & $-0.025$ & $-0.018$ \\
Women & Gradient Boosting & $-0.288$ & $-0.228$ & $-0.208$ & $-0.153$ & $-0.136$ \\
Women & Random Forest & $-0.183$ & $-0.216$ & $-0.249$ & $-0.247$ & $-0.284$ \\
\bottomrule
\end{tabular}
\end{adjustbox}
\caption{Temporal learning curves as a table. Entries at 20--80\% are means across 20 fixed-seed subsets of the 2023 training set; 100\% is the unique full-data fit. SVR is flat from 80\% to 100\%; RF/GB remain negative.}
\label{tab:learning_curves}
\end{table}

We also reran all three nonlinear models under four fixed outcome-trimming rules: $\pm20$, primary $\pm50$, and $\pm100$~s/day, plus 1.5-IQR bounds estimated from 2023 training outcomes only and then applied unchanged to 2024. Table~\ref{tab:outlier_sensitivity} reports the primary SVR learner. Estimates remain near zero under every rule; all Random Forest and Gradient Boosting estimates remain negative. Thus the null conclusion is not an artifact of the primary $\pm50$~s/day cutoff, although the exact point estimate is naturally sensitive near zero.

\begin{table}[htbp]
  \centering
  \caption{Outlier sensitivity for gender-separated SVR (train 2023 $\rightarrow$ test 2024). IQR bounds are estimated on training outcomes only.}
  \label{tab:outlier_sensitivity}
  \begin{tabular}{l rr rr}
    \toprule
    & \multicolumn{2}{c}{Men} & \multicolumn{2}{c}{Women} \\
    \cmidrule(lr){2-3} \cmidrule(lr){4-5}
    Rule & Test $n$ & $R^2$ & Test $n$ & $R^2$ \\
    \midrule
    $\pm20$ s/day & 1137 & $0.040$ & 654 & $0.025$ \\
    $\pm50$ s/day (primary) & 1164 & $0.044$ & 670 & $-0.018$ \\
    $\pm100$ s/day & 1167 & $0.034$ & 670 & $-0.018$ \\
    Training 1.5-IQR & 1033 & $-0.017$ & 557 & $0.011$ \\
    \bottomrule
  \end{tabular}
\end{table}

\subsection{Noise ceiling, permutation null, and classification robustness}
\label{app:null_diagnostics}

These three seed-controlled diagnostics test whether the RQ1 null reflects a lack of signal, a lack of data, or the intrinsic noisiness of the outcome.

\noindent\textbf{Reliability and noise ceiling:}
We decomposed variance in the Standardized race time within each athlete--season (one-way ICC; not detrended, so between-race variance includes systematic improvement: an approximation for the Galton overlay), and estimated the split-half reliability of the season improvement slope by splitting each athlete--season's races into odd/even sets, fitting a slope to each, correlating the two slopes across athlete--seasons, and applying the Spearman--Brown correction (Table~\ref{tab:noise_ceiling}). Individual race times are reliable (ICC $\approx0.8$), but the improvement slope, the object we try to predict, is only weakly reliable ($\approx0.23$--$0.28$). Because a construct cannot be predicted more accurately than it can be measured, this reliability is an approximate upper bound on any achievable test $R^2$, and it is close to zero. This bound is an approximation for a further reason: the odd/even split requires $\ge 4$ races per athlete--season, whereas the prediction models are evaluated on the full $\ge 2$-race sample ($n=1164$ men, $670$ women). Athletes with $4+$ races are a documented, more-persistent subgroup (Table~\ref{tab:retention}), so the reliability estimated on them need not transfer exactly to the broader sample; if anything, the shorter-schedule athletes excluded here have noisier slopes, which would lower the ceiling further rather than raise it.

\begin{table}[htbp]
\centering
\begin{tabular}{lrrr}
\toprule
Gender & Single-race ICC & Split-half $r$ & Reliability (S--B) \\
\midrule
Men & 0.78 & 0.13 & 0.23 \\
Women & 0.82 & 0.16 & 0.28 \\
\bottomrule
\end{tabular}
\caption{Reliability of the improvement outcome. Single-race ICC uses athlete--seasons with $\ge 2$ races; split-half reliability of the season slope uses athlete--seasons with $\ge 4$ races (odd/even split, Spearman--Brown corrected). The corrected reliability approximates the ceiling on any predictor's test $R^2$.}
\label{tab:noise_ceiling}
\end{table}

\noindent\textbf{Permutation null for held-out $R^2$:}
Shuffling the 2023 training outcome and refitting the compact-primary SVR $1{,}000$ times (seed 42), we compared the observed 2024 test $R^2$ to the permuted distribution. For men, observed $R^2=0.044$ versus a null mean of $-0.067$ (permutation $p=0.001$); for women, observed $R^2=-0.018$ versus a null mean of $-0.105$ ($p=0.001$). The observed values are reliably above chance, so the features carry a nonzero but practically negligible amount of information. Both still sit far below the reliability ceiling above.

\noindent\textbf{Classification of the tail:}
Because a continuous rate near zero may hide tail structure, we recast the task as predicting the top-quartile improver (threshold from the 2023 distribution) and report held-out AUC with 2{,}000-bootstrap CIs (Table~\ref{tab:classification_auc}). With all features, AUC reaches $0.79$--$0.81$; but removing features that are direct functions of the same race times used to build the first-to-last change score drops AUC to $0.66$--$0.69$. Most of the apparent tail skill is therefore the mechanical baseline$\to$change coupling when first time enters the change score, not novel predictive signal; a modest schedule-based residual remains.

\begin{table}[htbp]
\centering
\begin{adjustbox}{max width=\linewidth}
\begin{tabular}{lllr}
\toprule
Gender & Feature set & Classifier & AUC (95\% CI) \\
\midrule
Men & all features & Logistic & $0.81$ $[0.79,0.84]$ \\
Men & schedule-only & Logistic & $0.69$ $[0.66,0.72]$ \\
Women & all features & Logistic & $0.79$ $[0.75,0.82]$ \\
Women & schedule-only & Logistic & $0.66$ $[0.62,0.70]$ \\
\bottomrule
\end{tabular}
\end{adjustbox}
\caption{Held-out AUC for predicting the top-quartile improver (train 2023, test 2024). ``schedule-only'' removes baseline-coupled time features (first/best/worst/mean time, variability, slope, starting percentile). Random Forest results are similar and slightly lower.}
\label{tab:classification_auc}
\end{table}

\subsection{Mixed-effects explanatory models}
\label{app:rq1_mixed}

As an interpretable complement to the random forest, we fit linear mixed models with a random intercept for athlete. Fixed effects are $z$-scored \texttt{num\_races}, its square (\texttt{num\_races\textasciicircum2}), \texttt{season\_duration}, \texttt{starting\_percentile}, and calendar \texttt{year}, estimated separately within each sex. Coefficients are in seconds per day per one standard deviation increase in the predictor (outcome sign convention: more negative $=$ faster). Table~\ref{tab:mixedlm-rq1} lists estimates with 95\% confidence intervals.

The quadratic race-count term matches the non-monotone descriptive dose--response (Table~\ref{tab:dose_response}). For men the linear term is $-0.548$ ($p<0.001$) and the quadratic $0.170$ ($p=0.048$), with a vertex near the sparse $5+$ bin: some evidence of diminishing returns, not a confirmed inverted-U. The large starting-percentile coefficient ($-2.17$\,s/day per SD) is mechanically inflated by the same Galton coupling as binning on $t_1$. For women the curvature is imprecise. Year effects are sizable (men $-0.95$\,s/day per SD), consistent with residual calendar structure and the mean-shift sensitivity in the main text. Associational only.

\begin{table}[htbp]
  \centering
  \caption{Mixed-effects (MixedLM) fixed effects: estimate (seconds per day per 1 SD predictor), 95\% CI, two-sided $p$-value. Random intercept: athlete.}
  \label{tab:mixedlm-rq1}
  \begin{adjustbox}{max width=\linewidth}
  \begin{tabular}{l rrr rrr}
    \toprule
    & \multicolumn{3}{c}{Men} & \multicolumn{3}{c}{Women} \\
    \cmidrule(lr){2-4} \cmidrule(lr){5-7}
    Term & Est.\ & 95\% CI & $p$ & Est.\ & 95\% CI & $p$ \\
    \midrule
    Intercept
      & $-1.015$ & $[-1.290,\,-0.739]$ & $<0.001$
      & $-1.010$ & $[-1.381,\,-0.638]$ & $<0.001$ \\
    $z$\texttt{\_num\_races}
      & $-0.548$ & $[-0.856,\,-0.240]$ & $<0.001$
      & $-0.375$ & $[-0.794,\ \ 0.045]$ & $0.080$ \\
    $z$\texttt{\_num\_races\textasciicircum2}
      & $0.170$ & $[\ \ 0.001,\ \ 0.338]$ & $0.048$
      & $0.092$ & $[-0.144,\ \ 0.327]$ & $0.446$ \\
    $z$\texttt{\_season\_duration}
      & $0.173$ & $[-0.080,\ \ 0.427]$ & $0.180$
      & $0.219$ & $[-0.113,\ \ 0.551]$ & $0.196$ \\
    $z$\texttt{\_starting\_percentile}
      & $-2.173$ & $[-2.396,\,-1.950]$ & $<0.001$
      & $-1.724$ & $[-2.017,\,-1.430]$ & $<0.001$ \\
    $z$\texttt{\_year}
      & $-0.953$ & $[-1.161,\,-0.746]$ & $<0.001$
      & $-0.402$ & $[-0.669,\,-0.135]$ & $0.003$ \\
    \bottomrule
  \end{tabular}
  \end{adjustbox}
\end{table}

\noindent\textbf{Joint gender-interaction model:}
To test formally whether the race-count association differs by sex (rather than comparing the two separately fit columns above), we fit one random-intercept model on both sexes with female as the reference and male$\times$race-count interactions (Table~\ref{tab:mixedlm-interaction}). The main (female) linear term is negative and the curvature positive, as in the sex-specific fits, but both male interaction offsets are small and non-significant. That non-significance is not equivalence: the female sample is smaller, and the Wald 80\% minimum detectable linear interaction is $\approx 0.63$\,s/day per SD (from the CI width), larger than the point estimate, so a modest sex difference would likely be missed (Section~\ref{sec:rq3}).

\begin{table}[htbp]
\centering
\begin{adjustbox}{max width=\linewidth}
\begin{tabular}{lrrr}
\toprule
Term & Est.\ & 95\% CI & $p$ \\
\midrule
$z$\texttt{\_num\_races} (female) & $-0.389$ & $[-0.767,-0.011]$ & $0.044$ \\
$z$\texttt{\_num\_races\textasciicircum2} (female) & $0.091$ & $[-0.150,\ 0.332]$ & $0.458$ \\
male $\times$ $z$\texttt{\_num\_races} & $-0.157$ & $[-0.601,\ 0.287]$ & $0.487$ \\
male $\times$ $z$\texttt{\_num\_races\textasciicircum2} & $0.078$ & $[-0.212,\ 0.369]$ & $0.596$ \\
male (intercept offset) & $-0.091$ & $[-0.556,\ 0.373]$ & $0.700$ \\
\bottomrule
\end{tabular}
\end{adjustbox}
\caption{Joint mixed model with gender$\times$race-count interactions (female reference; random intercept per athlete; season duration, starting percentile, and year also included). Non-significant male interactions are compatible with underpower (approximate 80\% MDE $0.63$\,s/day per SD on the linear term), not with a demonstrated shared slope.}
\label{tab:mixedlm-interaction}
\end{table}

\section{Team selection robustness}
\label{app:team_selection}

Cross-sectional team models (Table~\ref{tab:team_robust}) leave open whether stronger programs both race more and place higher. We therefore run three checks. First, we form consecutive-season first differences within each (\texttt{team\_id}, gender) and regress $\Delta$top-15, and $\Delta$rank when both years have a top-25 finish, on $\Delta$max race count, $\Delta$depth ($\#$ athletes with $\ge 3$ starts), $\Delta$ERO, and $\Delta$roster size. Second, we refit cluster-robust GEE logistics of top-15 on $z$-scored max races and depth after adding $z$-scored roster size ($n$ athletes). Third, we define non-top-7 depth as the count of athletes outside the season-best top~7 with $\ge 3$ starts.

Among $328$ consecutive pairs, HC1 OLS coefficients of $\Delta$max races, $\Delta$depth, $\Delta$ERO, and $\Delta$roster size on $\Delta$top-15 are all non-significant ($p=0.61$, $0.45$, $0.090$, $0.098$). A Mundlak within--between LPM attributes the cross-sectional association to the between component ($+9.1$\,pp per max race $[5.3,12.9]$) with a null within component ($-0.2$\,pp $[-3.5,3.0]$). In percentage points, first-difference $\Delta$max races has slope $-0.87$\,pp per race $[-4.22,\,2.49]$ versus a cross-sectional $7.0$\,pp per race $[4.8,\,9.1]$; TOST at $\pm 5$\,pp per race gives $p=0.008$, supporting equivalence at that margin. First differences attenuate slopes through measurement error in $\Delta x$. That gap between within-team and cross-sectional max-race slopes is evidence against a simple causal reading for max races. Roster-size adjustment cuts the GEE OR for $z$(max races) from $2.56$ to $1.69$ $[1.05,2.71]$; depth collapses with size.

\begin{figure}[htbp]
\centering
\includegraphics[width=0.95\linewidth]{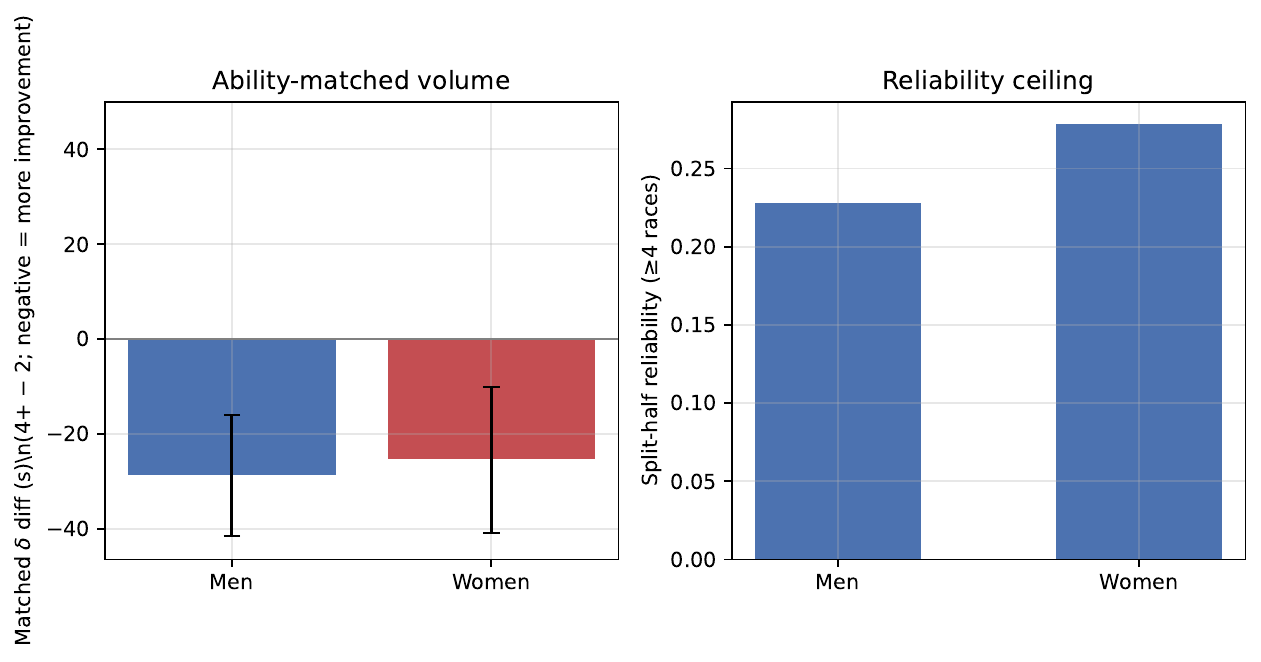}
\caption{Robustness-check summary: ability-matched volume contrasts on signed $\delta=t_L-t_1$ (negative $=$ more improvement for 4+; bootstrap CIs of the mean) and the split-half reliability ceiling on $\ge 4$-race seasons (the conservative $\ge 2$ ceiling equals that value by construction).}
\label{fig:robustness_checks}
\end{figure}

\noindent\textbf{Ability match, travel and size proxies, reliability:}
Figure~\ref{fig:robustness_checks} summarizes additional robustness and sensitivity checks.
\begin{itemize}
\item 1:1 matching on first Standardized time (caliper $0.25$\,SD) yields pooled 4+ vs.\ 2-race mean $\delta$ contrasts of about $-29$\,s (men, $n=527$; bootstrap 95\% CI on the mean $[-41,\,-16]$) and $-25$\,s (women, $n=303$; $[-41,\,-10]$) (negative $=$ more improvement for 4+). Wilcoxon tests on ranks are significant, and the mean CIs exclude zero.
\item GEE models with roster size and mean meet travel cut the max-race OR from $2.56$ to $1.67$; travel is null; within-team demeaned LPMs leave only size.
\item Across-season test--retest of the improvement rate is $r\approx0.05$, so rates barely predict themselves. The conservative $R^2$ ceiling for the $\ge 2$-race sample remains the split-half $\ge 4$-race reliability.
\item We do not encode historical-era race counts as experience: $43.9\%$ of historical athlete--years have only one recorded race and weather metadata are sparse, so counts systematically understate true volume.
\end{itemize}

\FloatBarrier
\section{Formulas: Standardization, field $\alpha$, and race prediction}
\label{app:field_em}
\label{app:std_formulas}

This appendix collects the times and estimators used in Sections~\ref{sec:data_standardization} and~\ref{sec:field_relative}. The club field factor is a LACCTiC-motivated analogue \citep{mazaheri2026lacctic}, not a bit-for-bit reimplementation (no track-PR initialization). Standardized coefficients are taken from the NRCD resource paper via \texttt{nrcd} \citep{karr2026nrcd,nrcdpackage}, not re-estimated here.

\noindent\textbf{Finish-time pairing (moved from main Method).}
Athlete fitness $f_i$ is held constant within the EM season window, so within-season fitness gains can load into $\alpha_j$ and make late (colder) meets look easier. The same EM is run after Converted Only, Standardized, and a homemade residual finish time (Equation~\ref{eq:temp_dew_residual}). We report $\rho(1/\alpha,T)$, the partial correlation controlling for day of year, and $\rho(1/\alpha,\mathrm{doy})$ directly. We also fit a joint EM on Converted times with temperature, dew point, and day-of-year covariates (``joint\_tdoy''). Next-raw MAE (Equation~\ref{eq:next_raw}) is reported for completeness but does not select the primary finish-time scale.

\subsection*{A.\ Converted Only and piecewise Standardized times}

\noindent\textbf{Distance (Converted Only).}
For a raw finish $t^{\mathrm{raw}}$ on measured distance $d$ with gender-specific target $d_\star\in\{8000,6000\}$\,m and Riegel exponent $b$ ($b=1.055$ men / $1.08$ women; package defaults, not fit on this sample),
\begin{equation}
t^{\mathrm{conv}}
=
t^{\mathrm{raw}}
\Bigl(\frac{d_\star}{d}\Bigr)^{b}.
\label{eq:riegel}
\end{equation}

\noindent\textbf{Metadata factors (missing inputs $\Rightarrow$ factor $=1$).}
Let $T,D$ be temperature and dew point ($^\circ$F), $H=T+D$, $G_+,G_-$ elevation gain and loss, and $z$ meet altitude. Define availability indicators $\mathbf{1}_{TD}=\mathbf{1}\{T,D\text{ observed}\}$, $\mathbf{1}_{G}=\mathbf{1}\{G_+,G_-\text{ both observed}\}$, $\mathbf{1}_{z}=\mathbf{1}\{z\text{ observed}\}$. With adopted constant $k=0.0016$ and threshold $H_0=100$ \cite{MaxPerformanceRunning},
\begin{equation}
\begin{aligned}
s(H)
&=
\begin{cases}
0, & \text{if }\mathbf{1}_{TD}=0\text{ or }H\le H_0,\\[2pt]
\dfrac{k}{100}(H-H_0)^{2}, & \text{if }\mathbf{1}_{TD}=1\text{ and }H>H_0,
\end{cases}
\\[3pt]
w(H)&=\dfrac{1}{1+s(H)},\\[6pt]
g(G_+,G_-)
&=
\begin{cases}
1, & \text{if }\mathbf{1}_{G}=0,\\[2pt]
(1.04)^{g_{\%}}\,(0.9633)^{\ell_{\%}}, & \text{if }\mathbf{1}_{G}=1,
\end{cases}
\end{aligned}
\label{eq:wg_defs}
\end{equation}
where $g_{\%},\ell_{\%}$ are percent grades from gain/loss over course length (Maurer-style defaults in \texttt{nrcd}). The altitude factor satisfies $a(z)=1$ whenever $\mathbf{1}_{z}=0$, and otherwise is the package meet-altitude correction (documented in \cite{karr2026nrcd}; not re-derived here). The Standardized time is the product
\begin{equation}
t^{\mathrm{std}}
=
t^{\mathrm{conv}}
\cdot w(H)
\cdot g(G_+,G_-)
\cdot a(z),
\label{eq:std_pieces}
\end{equation}
with each of $w,g,a$ already equal to $1$ when its inputs are missing (Equation~\ref{eq:wg_defs}).
On the comprehensive-era analysis export, $\mathbf{1}_{TD}\approx 99.9\%$, $\mathbf{1}_{z}\approx 100\%$, but $\mathbf{1}_{G}\approx 37\%$, so about $63\%$ of results are weather-adjusted without a grade factor. A percentage slowdown is corrected by division by $1+s$, not by subtracting the percentage from time. The adopted rule makes no adjustment for cold, wind, rain, mud, or footing; these can remain in residual meet difficulty. At $H=120,130,140,150$ the rule implies about $0.6\%$, $1.4\%$, $2.6\%$, $4.0\%$ slowdown.

\noindent\textbf{Environment residual and weather-path identity.}
Define $a:=t^{\mathrm{conv}}-t^{\mathrm{std}}$ and signed first-to-last change $\delta=t_L-t_1$ (negative $=$ faster). Equation~\eqref{eq:weather_path} is then an algebraic identity: $\delta^{\mathrm{conv}}-\delta^{\mathrm{std}}=a_L-a_1$, so the Converted--Standardized inflation magnitude is $E[a_1-a_L]$. Nulling elevation/grade/altitude inputs isolates the weather piece $a_{\mathrm{wx}}$ from the grade/altitude piece $a_{\mathrm{grade}}$ (Section~\ref{sec:weather_inflation}).

\subsection*{B.\ Field-factor EM (club LACCTiC analogue)}

Fix a gender and a preconditioned time $t^{(c)}\in\{t^{\mathrm{raw}},t^{\mathrm{conv}},t^{\mathrm{std}}\}$. Let $\mathcal{I}_j$ be athletes in meet $j$ with $|\mathcal{I}_j|\ge 5$, and $\mathcal{J}_i$ the meets of athlete $i$ with $|\mathcal{J}_i|\ge 2$. The club EM seeks $(f_i,\alpha_j)$ with $\alpha_j>0$, $f_i>0$, and scale pin $\operatorname{median}_j\alpha_j=1$, via damped alternating medians
\begin{equation}
\begin{aligned}
\alpha_j^{(m+1)}
&=
(1-\lambda)\,\alpha_j^{(m)}
+
\lambda\,
\operatorname{median}_{i\in\mathcal{I}_j}\frac{f_i^{(m)}}{t^{(c)}_{ij}},
\\
f_i^{(m+1)}
&=
(1-\lambda)\,f_i^{(m)}
+
\lambda\,
\operatorname{median}_{j\in\mathcal{J}_i}\bigl(\alpha_j^{(m+1)} t^{(c)}_{ij}\bigr),
\end{aligned}
\label{eq:field_em}
\end{equation}
with damping $\lambda=0.5$, ratio clips $\alpha_j\leftarrow\operatorname{clip}(\alpha_j,[0.5\alpha_j^{\mathrm{prev}},2\alpha_j^{\mathrm{prev}}])$, and iteration to relative tolerance $5\times 10^{-4}$ (max $300$ steps). Harder/slower meets have larger $t^{(c)}$ and smaller $\alpha$; field-adjusted time is $\tilde t_{ij}=\alpha_j t^{(c)}_{ij}$; leftover hardness is $h_j=1/\alpha_j$. Equation~\eqref{eq:field_em} never takes weather or elevation as covariates: those enter only through the choice of $t^{(c)}$.

\noindent\textbf{Isolation check.}
After fitting on $t^{(c)}$, report $\rho(h,T)$, $\rho(h,T\mid\mathrm{doy})$, and $\rho(h,\mathrm{doy})$. Near-zero leftover after $t^{(c)}=t^{\mathrm{std}}$ supports the weather heuristic; it is not a definitive validation of $k$.

\subsection*{C.\ Std-then-$\alpha$ hybrid and race prediction}

\noindent\textbf{Std-then-$\alpha$.}
Fit \eqref{eq:field_em} on $t^{\mathrm{std}}$ to obtain $(f_i,\alpha^{\mathrm{std}}_j)$. For athlete $i$'s ordered races $1,\ldots,L$ and target index $k\ge 2$, define prior ability and the predicted Standardized time at meet $j(k)$ by
\begin{equation}
\bar f_{ik}
=
\frac{1}{k-1}\sum_{\ell=1}^{k-1}
\alpha^{\mathrm{std}}_{j(\ell)}\,t^{\mathrm{std}}_{i\ell},
\qquad
\widehat t^{\mathrm{std}}_{ik}
=
\frac{\bar f_{ik}}{\alpha^{\mathrm{std}}_{j(k)}}.
\label{eq:std_then_alpha}
\end{equation}
Because $t^{\mathrm{std}}$ already includes whatever of $w,g,a$ are available, $\alpha^{\mathrm{std}}$ absorbs leftover field structure (mud, layout, and unmeasured grade when $\mathbf{1}_{G}=0$). Pure LACCTiC is the special case $t^{(c)}=t^{\mathrm{raw}}$; Converted-then-$\alpha$ uses $t^{(c)}=t^{\mathrm{conv}}$.

\noindent\textbf{Next-raw time and place (retrospective benchmark).}
Map back to the raw scale with the meet's mean raw-to-preconditioned ratio. For the hybrid $t^{(c)}=t^{\mathrm{std}}$,
\begin{equation}
\hat t^{\mathrm{raw}}_{ik}
=
\frac{\bar f_{ik}}{\alpha^{\mathrm{std}}_{j(k)}}\,
\overline{\!\left(\frac{t^{\mathrm{raw}}}{t^{\mathrm{std}}}\right)\!}_{j(k)},
\label{eq:next_raw}
\end{equation}
where $\bar f_{ik}$ uses athlete $i$'s races $1,\ldots,k-1$ only (Equation~\ref{eq:std_then_alpha}). In the present implementation, however, $\alpha^{\mathrm{std}}_{j(k)}$ is estimated by the full-season EM and the meet scale is computed from target-meet results. Both therefore use information unavailable before race $k$, including the target meet's realized times. The resulting score is a retrospective next-race reconstruction benchmark, not a deployable pre-race forecast. Reconstructed place among the set $\mathcal{F}_{j(k)}$ is
\begin{equation}
\widehat{\mathrm{place}}_{ik}
=
1+\sum_{i'\in\mathcal{F}_{j(k)}\setminus\{i\}}
\mathbf{1}\!\bigl\{\hat t^{\mathrm{raw}}_{i'k}<\hat t^{\mathrm{raw}}_{ik}\bigr\}
+
\tfrac12\sum_{i'\in\mathcal{F}_{j(k)}\setminus\{i\}}
\mathbf{1}\!\bigl\{\hat t^{\mathrm{raw}}_{i'k}=\hat t^{\mathrm{raw}}_{ik}\bigr\},
\label{eq:prerace}
\end{equation}
i.e., the ascending average-rank of $\hat t^{\mathrm{raw}}$ within the meet (ties split). Matched last-race MAE and within-meet place Spearman prefer Std-then-$\alpha$ over either ingredient alone as a retrospective benchmark (Figure~\ref{fig:std_lacctic_race}).

\noindent\textbf{Homemade residual finish time (negative control).}
Within sex,
\begin{equation}
\log t^{\mathrm{conv}}_{ir}-\overline{\log t}^{\mathrm{conv}}_{i}
=
\beta_T(T_{ir}-50)+\beta_D(D_{ir}-40)+\varepsilon_{ir},
\label{eq:temp_dew_ols}
\end{equation}
\begin{equation}
t^{\mathrm{tdew}}_{ir}
=
t^{\mathrm{conv}}_{ir}
\exp\bigl(-[\beta_T(T_{ir}-50)+\beta_D(D_{ir}-40)]\bigr),
\label{eq:temp_dew_residual}
\end{equation}
with missing $T$ or $D$ contributing zero (same rule as $w=1$). This is an in-sample negative control, not a competing standardization formula.

\noindent\textbf{joint\_tdoy (this paper).}
As a further in-sample check on Converted Only, replace the nonparametric EM with the linear score
\begin{equation}
\log t^{\mathrm{conv}}_{ij}
=
\log f_i
-
\log\alpha_j
-
\beta^\top x_{ij}
+
\varepsilon_{ij},
\qquad
x_{ij}=(T_{ij}-\bar T,\,D_{ij}-\bar D,\,\mathrm{doy}_{ij}-\overline{\mathrm{doy}}),
\label{eq:joint_tdoy}
\end{equation}
fit by alternating updates in $(f,\alpha,\beta)$. Estimated $\beta_T\approx 0.00064$ (men) / $0.00079$ (women) ($\approx 0.064$--$0.079\%$ per $^\circ$F); leftover $\rho(1/\alpha,T)\approx 0.13$--$0.14$; next-raw MAE $\approx 45$--$50$\,s. The linear form is smaller than the heuristic over the warm range and does not clear temperature the way Standardized-then-$\alpha$ does.

\section{Multiplicity Accounting}
\label{app:multiplicity}

We corrected for multiple comparisons \emph{within} families of related tests rather than across the entire paper, because the families address distinct questions and a single paper-wide correction would be both unconventional and needlessly conservative across unrelated hypotheses. For transparency, Table~\ref{tab:multiplicity} enumerates every family of inferential tests reported, its size, the correction applied, and whether any member survives. Point estimates with confidence intervals (e.g., the mixed-effects coefficients, reliability, and pooled/continuous team models) are reported as estimation rather than as null-hypothesis screening and are listed for completeness. Readers preferring a global adjustment can apply a Bonferroni factor of $\approx 5$ to the within-family $\alpha$ thresholds (not to $p$-values loosely). Under that stricter screen the reliability ceiling, weather-inflation contrasts, and pooled/continuous team association remain; exploratory per-cell dose-response and per-cell risk ratios do not.

\begin{table}[htbp]
\centering
\begin{adjustbox}{max width=\linewidth}
\begin{tabular}{p{4.4cm}rp{3.2cm}p{4.2cm}}
\toprule
Family & \# tests & Correction & Outcome \\
\midrule
Per-cell top-15 risk ratios & 6 & Bonferroni $\alpha=0.0083$ & None survives per-cell; pooled/continuous models significant \\
Dose--response pairwise contrasts & 6 & Bonferroni / BH $\alpha=0.05$ & Only men's 4-vs-2 survives (exploratory) \\
Starting-ability $4+$ vs $2$ strata & 8 & BH FDR $\alpha=0.05$ & 5/8 survive (Q3/Q4 both sexes; men Q1) \\
Model pairwise CV comparisons (RQ1/RQ2) & 15 per task & Bonferroni & 8/15 (RQ1) and 2/15 (RQ2) nominal; none survive \\
Permutation-$R^2$ tests (per sex) & 2 & report exact $p$ & Both $p=0.001$ \\
Roster-depth GEE \& mixedLM & estimation & CIs / exact $p$ & Depth OR reported with CIs \\
\bottomrule
\end{tabular}
\end{adjustbox}
\caption{Inferential-test families, sizes, and within-family multiplicity corrections. Estimation results (mixed-effects coefficients, reliability, pooled/cluster-robust team models, depth coefficients) are reported with CIs rather than corrected $p$-values.}
\label{tab:multiplicity}
\end{table}

\end{document}